\documentclass[useAMS,usenatbib]{mn2e}

\usepackage{graphicx,amssymb,hyperref}
\usepackage[fleqn]{amsmath}

\newcommand{\ellipticity}{\varepsilon}
\newcommand{\ei}{\bmath{\ellipticity}_{\rm int}}
\newcommand{\eg}{\bmath{\ellipticity}_{\rm gal}}
\newcommand{\ep}{\bmath{\ellipticity}_{\rm PSF}}
\newcommand{\ec}{\bmath{\ellipticity}_{\rm NC}}
\newcommand{\eo}{\bmath{\ellipticity}_{\rm obs}}

\newcommand{\weg}{\bmath{\ellipticity}_{\rm gal\,w}}
\newcommand{\wep}{\bmath{\ellipticity}_{\rm PSF\,w}}
\newcommand{\wec}{\bmath{\ellipticity}_{\rm NC\,w}}
\newcommand{\weo}{\bmath{\ellipticity}_{\rm obs\,w}}

\newcommand{\ri}{R_{\rm int}}
\newcommand{\rg}{R_{\rm gal}}
\newcommand{\rp}{R_{\rm PSF}}
\newcommand{\rc}{R_{\rm NC}}
\newcommand{\ro}{R_{\rm obs}}

\newcommand{\wrg}{R_{\rm gal\,w}}
\newcommand{\wrp}{R_{\rm PSF\,w}}

\newcommand{\add}{\mathcal{A}}
\newcommand{\mult}{\mathcal{M}}
\newcommand{\sigs}{\sigma_{\rm sys}}
\newcommand{\bc}{\bmath{c}}
\newcommand{\bd}{\bmath{\delta}}
\newcommand{\pg}{P_{\gamma}}
\newcommand{\peo}{P_{\eo}}
\newcommand{\pec}{P_{\ec}}
\newcommand{\pep}{P_{\ep}}
\newcommand{\pw}{P_{{\rm {w}}_\ellipticity}}
\newcommand{\pr}{P_{R}}

\newcommand{\pwep}{W_{\ep}}

\newcommand{\sigc}{\sigma^2_{\bc}}

\newcommand{\be}{\begin{equation}}
\newcommand{\ee}{\end{equation}}
\newcommand{\ba}{\begin{eqnarray}}
\newcommand{\ea}{\end{eqnarray}}
\newcommand{\nn}{\nonumber \\}

\def\app#1#2{%
  \mathrel{%
    \setbox0=\hbox{$#1\sim$}%
    \setbox2=\hbox{%
      \rlap{\hbox{$#1\propto$}}%
      \lower1.2\ht0\box0%
    }%
    \raise0.5\ht2\box2%
  }%
}

\def\mean#1{\left\langle#1\right\rangle}
\def\variance#1{\left\langle\left|#1\right|^2\right\rangle}

\def\simlt{\lower.5ex\hbox{$\; \buildrel < \over \sim \;$}}
\def\simgt{\lower.5ex\hbox{$\; \buildrel > \over \sim \;$}}

\def\etc{{\it etc}}
\def\ie{{\it i.e.}}
\def\egg{{\it e.g.}}
\def\cf{{\it c.f.}}
\usepackage{graphicx} 

\title[Origins \& requirements for weak lensing systematics]{
Origins of weak lensing systematics, and requirements on future instrumentation (or knowledge of instrumentation)}
\author[R.\ Massey et al.]{Richard Massey$^{1}$\thanks{e-mail: {\tt r.j.massey@durham.ac.uk}}, 
Henk Hoekstra$^{2}$\thanks{e-mail: {\tt hoekstra@strw.leidenuniv.nl}}, 
Thomas Kitching$^{3}$\thanks{e-mail: {\tt tdk@roe.ac.uk}}, Jason Rhodes$^{4,5}$, Mark \newauthor Cropper$^{6}$, 
J\'er\^ome Amiaux$^{7}$, David Harvey$^{3}$, Yannick Mellier$^{8,7}$, Massimo Meneghetti$^{9}$,\newauthor 
Lance Miller$^{10}$, St\'ephane Paulin-Henriksson$^{7}$, Sandrine Pires$^{7}$, Roberto Scaramella$^{11}$\newauthor and Tim Schrabback$^{12,13}$ \\
$^{1}$ Institute for Computational Cosmology, Durham University, South Road, Durham DH1 3LE, UK  \\ 
$^{2}$ Leiden Observatory, Leiden University, P.O.\ Box 9513, 2300 RA, Leiden, The Netherlands\\
$^{3}$ University of Edinburgh, Royal Observatory, Blackford Hill, Edinburgh EH9 3HJ, UK  \\
$^{4}$ Jet Propulsion Laboratory, California Institute of Technology, 4800 Oak Grove Drive, Pasadena, CA 91109, USA\\
$^{5}$ California Institute of Technology, 1201 E California Blvd., Pasadena, CA 91125, USA\\
$^{6}$ Mullard Space Science Laboratory, University College London, Holmbury St Mary, Dorking, Surrey RH5 6NT, UK\\
$^{7}$ Service d$\arcmin$Astrophysique, CEA Saclay, Gif sur Yvette, 91191, France\\
$^{8}$ Institut d$\arcmin$Astrophysique de Paris, UMR7095 CNRS, Universit\'e Pierre et Marie Curie, 98 bis Boulevard Arago, 75014 Paris, France\\
$^{9}$ INAF, Osservatorio Astronomico di Bologna, via Ranzani 1, 40127 Bologna, Italy\\
$^{10}$ Department of Physics, University of Oxford, The Denys Wilkinson Building, Keble Road, Oxford, OX1 3RH, UK\\
$^{11}$ INAF, Osservatorio Astronomico di Roma, via Frascati 33, 00040 Monteporzio Catone, Italy\\
$^{12}$ Kavli Institute for Particle Astrophysics and Cosmology, Stanford University, 382 Via Pueblo Mall, Stanford, CA 94305, USA\\
$^{13}$ Argelander-Institut f\"ur Astronomie, Auf dem H\"ugel 71, D-53121 Bonn, Germany \vspace{-5mm} 
}
\begin{document}
\date{ Accepted ---. Received ---; in original form \today.}

\pagerange{\pageref{firstpage}--\pageref{lastpage}} \pubyear{2011}

\maketitle

\label{firstpage}

\begin{abstract}

The first half of this paper explores the origin of systematic biases in the measurement of weak gravitational lensing.
Compared to previous work, we expand the investigation of PSF instability
and fold in for the first time the effects of non-idealities in electronic imaging detectors
and imperfect galaxy shape measurement algorithms.
Together, these now explain the additive $\add(\ell)$ and multiplicative $\mult(\ell)$ systematics typically reported in current lensing measurements.
We find that overall performance is driven by a product of a telescope/camera's {\em absolute performance}, and our {\em knowledge about its performance}.

The second half of this paper propagates any residual shear measurement biases through to their effect on cosmological parameter constraints.
Fully exploiting the statistical power of Stage~IV weak lensing surveys will require additive biases $\overline\add\simlt 1.8\times10^{-12}$ and multiplicative biases  $\overline\mult\simlt 4.0\times10^{-3}$.
These can be allocated between individual budgets in hardware, calibration data and software, using results from the first half of the paper.

If instrumentation is stable and well-calibrated, we find extant shear measurement software from GREAT10 
already meet requirements on galaxies detected at S/N=40. Averaging over a population of galaxies with a realistic distribution of sizes, it also meets requirements
for a 2D cosmic shear analysis from space.
If used on fainter galaxies or for 3D cosmic shear tomography, existing algorithms would need calibration on simulations to avoid introducing bias at a level similar to the statistical error.
Requirements on hardware and calibration data are discussed in more detail in a companion paper.
Our analysis is intentionally general, but is specifically being used to drive the hardware and ground segment performance budget for the design of the European Space Agency's recently-selected Euclid mission.

\end{abstract}

\begin{keywords}
gravitational lensing --- cosmology: cosmological parameter --- instrumentation: detectors --- methods: data analysis \vspace{-9mm}
\end{keywords}

\section{Introduction}

Statistical measurements of weak gravitational lensing in a large sample of galaxies offer a direct way to probe the dark sector of the Universe \citep[see reviews by][]{hrev,mrev}. Gravitational lensing is the deflection of light from distant galaxies during its journey to us, by an amount that depends on the intervening distribution of matter (including dark matter) and the geometry of spacetime (which is currently governed by dark energy). The deflection of light produces slight shear distortions in the galaxies' apparent shapes, and adjacent galaxies appear to line up in characteristic patterns across the sky.

Galaxy ellipticities are typically distorted only a few percent by weak gravitational lensing.
Detecting this tiny signal is difficult because the image shapes are also changed an order of magnitude more by convolution with the Point Spread Function (PSF) of the telescope, detector and atmosphere, as well as by distortion in the camera.
These other effects must be modelled and corrected; even subtle residual contributions can significantly bias cosmological measurements.

In the first half of this paper we explore three types of error that affect galaxy shape measurement:
\begin{itemize}\vspace{-2mm}
\item inaccuracies in the model of the convolutional PSF, from which observed galaxy shapes must be deconvolved
\citep[this builds upon work by][]{sph08}.
\item inaccuracies in correction for any effect that cannot be treated as a deconvolution.
This includes detector effects such as Charge Transfer Inefficiency in CCDs or Inter-Pixel Capacitance in HgCdTe devices, 
which perturb pixel values in a nonlinear fashion.
\item inaccuracies in the measurement of galaxy shapes. 
Minimising noise, particularly in faint galaxies, forces measurement methods to apply pixel weights which must subsequently be undone.
\end{itemize}\vspace{-2mm}
We propagate these measurement errors through a tomographic cosmic shear analysis
(theory developed by \citealt{hu99,jai03,ber04} and measurements obtained by \citealt{kit3d,cosmosm,sch10}) 
to determine the bias they induce upon constraints on the dark energy equation of state parameter $w$
\citep{tomo1,tomo2,tomo3}. 

In the second half of this paper, we establish requirements on additive and multiplicative cosmic shear systematics to meet future scientific goals. 
We also use our earlier results to consider how residual additive biases can be empirically identified and removed, and assess the impact of residual multiplicative biases that cannot be self-consistently identified within a data set.
Our analysis is intentionally performed with a scope sufficiently general to cover any future Stage~IV weak gravitational lensing survey.
It is particularly motivated by, and drives the hardware and ground segment performance budget for the design of the European Space Agency's recently-selected Euclid mission \citep{euclidredbook}.
This work generalises the conclusions of \citet{hardsoft}.
During the final preparation of this paper, \citet{cha12} posted to the arxiv an analysis of future prospects for the Large Synoptic Survey Telescope (LSST). 
There is some overlap in ambition, but complementary methodology.
Like Chang et al.\, we employ a bottom-up approach in this paper, propagating various instrumental imperfections through to errors on cosmological parameters.
However, rather than simulating the detailed performance of a baseline telescope model, we work analytically to build a general framework for propagating general system performance.
In a companion paper, \citet{cropper12}, this allows us to perform a top-down, systems engineering analysis: starting from the science requirements and flowing down to requirements on subsystem performances.
Using the understanding from this paper, the total error budget and mitigation can be sensibly allocated between individual budgets in hardware, calibration data and software performance.

This paper is organised as follows.
In Section~\ref{sec:defs}, we define the basic galaxy shape and cosmological quantities of interest that would be measured in a weak lensing experiment with no (or idealised) errors.
In Section~\ref{sec:problems}, we explore the various types of error that can be introduced during realistic galaxy shape measurement and may prevent recovery of the true signal. 
Our underlying approach builds upon the work of \citet{sph08} -- however the mathematical expressions rapidly lengthen when we introduce more sources of error. For clarity, we therefore choose to evolve the formalism in three stages, one for each source of error.
In Section~\ref{sec:cosmo}, we derive requirements on shear measurement biases for a cosmic shear survey seeking to measure dark energy. 
In Section~\ref{sec:met}, we determine whether those requirements are met by extant shear measurement software described in the literature.
We do this at fixed galaxy fluxes and, using our results from the first half of the paper, averaging over the full population of galaxies that will be seen by a survey.
We conclude in Section~\ref{sec:conc}.

\section{Idealised weak lensing measurement} \label{sec:defs}

\subsection{Perfect shear measurement} \label{sec:perfect}

Many techniques have been developed to precisely measure the shapes of galaxies \citep[see][]{great08,great10}.
For the sake of a concrete example, we shall consider the generic class of methods based upon galaxies' quadrupole moments.
In a method based on unweighted quadrupole moments \citep[see][]{bs01}, the shape of any localised object in a 2D image $I(r,\theta)$ can be quantified via 
its size 
\be \label{eqn:momdefr}
R^2\equiv\frac{\iint I(r,\theta)~r^2~r{\rm d}r{\rm d}\theta}{\iint I(r,\theta)~r{\rm d}r{\rm d}\theta}
\ee
and complex ellipticity
\be \label{eqn:momdefe}
\bmath{\ellipticity}= \ellipticity_1+ i\ellipticity_2 \equiv
\frac{\iint I~r^2{\rm e}^{2i\theta}~r{\rm d}r{\rm d}\theta}{\iint I~r^2~r{\rm d}r{\rm d}\theta}.
\ee

Gravitational lensing magnifies and shears a galaxy of intrinsic size $\ri$ and ellipticity $\ei$ into one of size $\rg>0$ and ellipticity
\be
\eg=\ei+\pg\bmath{\gamma},
\ee
where the shear `polarizability' $\pg\equiv\partial\ei/\partial\gamma\approx1.86$ is the amount by which the ellipticity of a galaxy changes during gravitational lensing\footnote{Strictly, $\pg$ depends on galaxy morphology and
is also a $2\times2$ tensor acting separately on the real and imaginary components of ellipticity. On average, however, it is very close to the identity tensor times a real scalar $2-\langle |\ei|^2\rangle$ \citep{rrg}. \citet{lea07} show that $\langle |\ei|^2\rangle$ is consistent with a constant value of $2\times0.26^2$ for galaxies to at least redshift $z=2.6$. We greatly simplify subsequent analysis by assuming scalar $\pg\approx1.86$.\label{foot:pg}} \citep{ksb,lk97}. 
When this galaxy is imaged by any camera, it is convolved with a Point Spread Function (PSF, of size $\rp$ and ellipticity $\ep$), producing\footnote{Relationships \eqref{eqn:rodef_sph} and \eqref{eqn:eodef_sph} are exact using unweighted moments, but hold for some other methods only if both the galaxy and the PSF are approximately Gaussian. We shall return to this issue in Section~\ref{sec:step}.} an observed source of larger size 
\be \label{eqn:rodef_sph}
\ro^2=\rg^2+\rp^2
\ee
and perturbed ellipticity
\be \label{eqn:eodef_sph}
\eo=\eg+\frac{\rp^2}{\rg^2+\rp^2}\left(\ep-\eg\right).
\ee

Weak lensing analyses observe the shape of each galaxy then try to correct it for (or deconvolve it from) the PSF, 
to recover the galaxy's true ellipticity. 
The system PSF can be measured from stars also within the field of view.
Rearranging equations~\eqref{eqn:rodef_sph} and \eqref{eqn:eodef_sph} so that only observable quantities appear on the right hand side, the galaxy's ellipticity is
\be \label{eqn:egdef_sph}
\eg=\frac{\eo\ro^2-\ep\rp^2}{\ro^2-\rp^2}~.
\ee
The intrinsic ellipticity of individual galaxies is uninteresting so, to isolate the cosmologically relevant information, ellipticity is normalised into a shear estimator
\be \label{eqn:pgdef1}
\bmath{\widehat\gamma}\equiv(\pg)^{-1}~\eg~.
\ee
This ensures that, averaging over a large number of galaxies,
\be
\langle\bmath{\widehat\gamma}\rangle=(\pg)^{-1}\langle\ei\rangle+(\pg)^{-1}\pg\,\langle\bmath{\gamma}\rangle \\
\ee
and we recover $\langle\bmath{\widehat\gamma}\rangle=\langle\bmath{\gamma}\rangle$ so long as the intrinsic galaxy ellipticities are random and hence $\mean{\ei}=0$ \citep[but see][for instances of `intrinsic alignments' when this does not hold]{ia1,ia2,ia3,ia4,ia5,ia6}.

The average shear is zero so, to compare to theoretical models, the measured shears are then combined into two-point correlation functions 
\ba \label{eqn:cijdef}
\xi_+(\theta,z_A,z_B)&\equiv&\left\langle{\bmath{\gamma}}_A{\bmath{\gamma}}_B^*\right\rangle(\theta,z_A,z_B),\\
\xi_-(\theta,z_A,z_B) &\equiv&{\rm Re}\big(\left\langle{\bmath{\gamma}}_A{\bmath{\gamma}}_B\right\rangle(\theta,z_A,z_B)\big),
\ea
where the angle brackets indicate averaging over all pairs of galaxies $A$ and $B$ in a survey that are at redshifts $z_A$ and $z_B$ and separated on the sky by an angle $\theta$,
or within bins around those values \citep{cri00,bs01}.
The correlation functions trace a cosmological, `cosmic shear' signal at $\theta>0$.
Results are often expressed in terms of the shear power spectrum $C(\ell)$, the Fourier transform of a weighted sum of $\xi_\pm(\theta)$.

If galaxy shapes are autocorrelated with themselves, a zero-lag term
$\sigma^2_{\bmath{\gamma}}\delta(\theta$=0) is added. 
This is included by \citet[][equation~11]{sph08} but we disregard it
because it can be readily avoided by excluding such galaxy pairs in practice.
If the $\sigma^2_{\bmath{\gamma}}$ autocorrelation term were not removed from an analysis, it would be white noise, independent of scale in the Fourier transform. 
This must be marginalised over as an unknown constant of integration, subtracted from measurements, or added to theoretical models.

\subsection{Parametric shear measurement bias} \label{sec:stepform}

Deviations from perfect shear measurement are commonly parameterised following the Shear TEsting Programme \citep[STEP;][]{step1,step2} as
\be \label{eqn:stepgamma}
\widehat{\bmath{\gamma}}=(1+m)\bmath{\gamma}+\bc .
\ee
We shall henceforth represent all real-world, imperfect measurements using a hat. 

\subsubsection{Constant shear measurement bias} \label{sec:stepformc}

We first consider shear measurements that have small additive bias $\bc $ with constant mean $\langle \bc \rangle$ and random noise $\sigma_{\bc}$, plus small multiplicative bias $m$ with constant mean $\mean{m}$ and random noise $\sigma_m$.
Pairs of these shear measurements can be folded through the calculation of a correlation function \eqref{eqn:cijdef} to produce 
\ba \label{eqn:chat}
\widehat{\xi_+}(\theta,z_A,z_B)
 \equiv\left\langle\widehat{\bmath{\gamma}}_A\widehat{\bmath{\gamma}}_B^*\right\rangle~~~~~~~~~~~~~~
 ~~~~~~~~~~~~~\\
 =\big\langle(1+m)(1+m)\big\rangle \,\xi_++\variance{\bc}
\ea
plus cross terms only in the presence of shear-dependent selection effects \citep[see \egg][and the discussion in Appendix~A]{jain06}.

Taking the Fourier transform to yields a power spectrum spectrum\footnote{\citet[][eqn.~13]{ar08} rearrange \eqref{eqn:chat} as a Taylor series expansion of the measured correlation function
\be \label{eqn:chatar}
\widehat{C}(\ell)\equiv C(\ell)+\Big\{\add_0+\add_1\,C(\ell)+\dots\Big\} \nonumber
\ee
and label everything inside the curly brackets as different types of `additive error' $C_\ell^\mathrm{sys}$.
Simple multiplicative biases easily arise, so we instead find it helpful to keep terms $\add$ and $\mult$ separate.
Because of the shape of the $\Lambda CDM$ cosmological power spectrum, they are nearly orthogonal
and have quite different implications (see Section~\ref{sec:cosmo}).
We therefore restrict our notation for $\add$ to refer solely to pure additive terms.}
\be \label{eqn:chat0}
\widehat{C}(\ell,z_A,z_B) = \left(1+\mult\right)C(\ell,z_A,z_B)+\add,
\ee
where
\be \label{eqn:aeq0}
\add=0
\ee
\ba \label{eqn:Mapprox}
\mult=2\langle m\rangle+\langle m^2\rangle~=~2\langle m\rangle+\langle m\rangle^2+\sigma^2_m\,\delta(0) \\
\approx 2\langle m\rangle \,. ~~~~~~~~~~~~~~~~~~~~~~~~~~~~~~~~~~~~~~~~~~~~\, \label{eqn:Mapprox2m}
\ea
We have expanded the mean squared error $\mean{m^2}$ term but note that $\mean{m}^2\ll\mean{m}$ and that $\mult$ terms arise only from the correlation of galaxies with other galaxies. Thus a constant multiplicative bias in shear measurements leads to a similarly constant multiplicative bias in a measurements of the shear power spectrum.
If the autocorrelation terms discussed in Section~\ref{sec:perfect}) are included, equation~\eqref{eqn:aeq0} gains an additional white noise term
\be \label{eqn:siggamstep}
\sigma^2_{\bmath{\gamma}} =  \left[ 1+2\langle m\rangle+\langle m^2\rangle \right] \pg^{-2}\sigma^2_{\eg} +\sigma_{\bc}^2\,.
\ee
This notation is also discussed in \citet{g10res}.


\subsubsection{Spatially/temporally varying shear measurement bias} \label{sec:stepformv}

We next consider shear measurement biases in which $\bc$ and $m$ vary from galaxy to galaxy, and their deviations from a mean value can be correlated in patterns across a survey.
In general, $\add$ and $\mult$ become functions of scale, orientation on the sky, and redshift
\ba \label{eqn:chat2}
\widehat{C}(\bmath{\ell},z_A,z_B) =\sum_{\bmath{\ell}'}\big(1+\mult(\bmath{\ell},\bmath{\ell}^\prime,z_A,z_B)\big)C(\bmath{\ell}^\prime,z_A,z_B)~~~~\nn +\add(\bmath{\ell},z_A,z_B)
\ea
\citep[see Appendix A of][]{g10res}. 
The additive systematics $\add$ now include a contribution from the spatially varying additive shear measurement bias.
The $\mult$ matrices mix power from different scales, as well as physical $E$-mode and non-physical $B$-mode signals, where $C=C_E+iC_B$. 
Anisotropic errors could arise from PSF terms in off-axis cameras, from non-square pixels, from some detector effects, or in ground-based surveys where gravity loading and the prevailing wind can impose preferred directions. 
In this paper, we shall only consider the simpler situation in which the systematic errors are isotropic on average within a survey. In this case, the matrices are diagonal, so $\add$ and $\mult$ become functions of only scale and redshift
\be \label{eqn:chat3}
\widehat{C}(\ell,z_A,z_B) = \big(1+\mult(\ell,z_A,z_B)\big)C(\ell,z_A,z_B)+\add(\ell,z_A,z_B).
\ee
Using the notation $\sigma^2[x]$ to represent the covariance about the mean of error $\delta x$ in pairs of galaxies separated by $\theta>0$, we find
\be \label{eqn:Aapproxvar}
\add(\ell,z_A,z_B) = \sigma^2[|\bc|](\ell,z_A,z_B) 
\ee
\be \label{eqn:Mapproxvar}
\mult(\ell,z_A,z_B) \approx \sigma^2[{m}](\ell,z_A,z_B) + 2\langle m\rangle(z_A,z_B).
\ee
\citet{sph08} miss the second half of equation~\eqref{eqn:Mapproxvar}
because they ignore bias terms when expanding mean squared errors. 
This was reasonable for purely additive systematics, as spatially constant terms disappear during a Fourier transform; but in this case we judge that the bias term is likely to be the most problematic.
Instead, we find that cosmic shear biases arise from a combination of (a) absolute biases in shear measurement and (b) uncertainty in or lack of knowledge about shear measurements. 
This dichotomy will emerge as a general result throughout Section~\ref{sec:problems}.

\section{Realistic weak lensing measurement errors} \label{sec:problems}

\subsection{Imperfect PSF correction} \label{sec:sph}

Errors in shear measurement can arise from several sources. 
For example, our model of the PSF will inevitably be imperfect because it is obtained from noisy stars and must be interpolated to the position and colour of each galaxy \citep[\egg][]{hoe04,man05,jain06,sparsity,ed09}. 

Via a first order Taylor series expansion of equation~\eqref{eqn:egdef_sph},
model errors in the PSF size $\delta(\rp^2)$ and ellipticity $\bd\ep$ propagate into an imperfect estimate of the galaxy ellipticity
\be \label{eqn:eghatdef1}
\widehat{\eg}\approx\eg + \frac{\partial\eg}{\partial(\rp^2)}\delta(\rp^2) + \frac{\partial\eg}{\partial\ep}\delta\ep.
\ee
The partial derivatives of \eqref{eqn:egdef_sph} are
\be \label{eqn:partial1}
\frac{\partial\eg}{\partial(\rp^2)} = \frac{\ro^2}{\left(\ro^2-\rp^2\right)^2}(\eo-\ep) = \frac{\eg-\ep}{\rg^2}~,
\ee
\be \label{eqn:partial2}
\frac{\partial\eg}{\partial\ep} = -\frac{\rp^2}{\ro^2-\rp^2} = -\frac{\rp^2}{\rg^2} ~,
\ee
and the derivative with respect to the other real/imaginary component of the PSF ellipticity is zero.
In equations~\eqref{eqn:partial1} and \eqref{eqn:partial2}, the first equality is expressed in terms that are observable in an image, and the second equality reflects fundamental source properties. Inserting the latter into \eqref{eqn:eghatdef1} yields
\ba \label{eqn:eghatdef2}
\widehat{\eg}\approx\left\{1+\frac{\delta(\rp^2)}{\rg^2}\right\}\eg  - ~~~~~~~~~~~~\nn
      \left\{\frac{\rp^2}{\rg^2}\bd\ep + \frac{\delta(\rp^2)}{\rg^2}\ep \right\}.
\ea
Arranged thus (\cf\ \citealt{sph08} eqn.~8), the last two terms display an elegant symmetry: 
the product of the PSF size and our knowledge of its ellipticity, 
then its absolute ellipticity and our knowledge of its size.
The STEP parameters can be easily read off from this expression.
Note that if the PSF ellipticity is known perfectly ($\bd\ep=\bmath{0}$), $\bc =-m\ep/\pg$ and the two are related.

When folding this imperfect shear estimator through the calculation of a correlation function \eqref{eqn:cijdef}, to multiply out some angle brackets we follow \citet{sph08} in assuming that 
inaccuracies in the model of the PSF shape are independent of the shape of the PSF and the size of galaxies to which it is applied. 
If that does not hold, the angle brackets cannot be separated and some cross-terms can be introduced that are computed in Appendix~A. 
An additional assumption that \citet{sph08} and we make is that the size of the PSF is roughly constant across the survey, such that $\mean{\rp^2\rp^2}(\ell)\approx\mean{\rp^4}$.
In exact correspondence to the various terms of equations~\eqref{eqn:Aapproxvar} and \eqref{eqn:Mapproxvar}, we find
\ba \label{eqn:sigsdef2}
\add(\ell,z_A,z_B)=\frac{1}{\pg^{~2}} \left\langle\frac{\rp^4}{\rg^4}\right\rangle \variance{\bd\ep} (\ell,z_A,z_B)  \nn
         + \frac{\variance{\ep}}{\pg^{~2}} \left\langle\frac{\rp^4}{\rg^4}\right\rangle \frac{\variance{\delta(\rp^2)}}{\mean{\rp^4}} (\ell,z_A,z_B) 
\ea
\ba \label{eqn:mdef}
\mult(\ell,z_A,z_B) = \left\langle\frac{\rp^4}{\rg^4}\right\rangle \frac{\variance{\delta(\rp^2)}}{\mean{\rp^4}}(\ell,z_A,z_B)~~ \nn
                                  + 2 \left\langle\frac{\rp^2}{\rg^2}\right\rangle \frac{\mean{\delta(\rp^2)}}{\mean{\rp^2}}(z_A,z_B) \,.
\ea
In combination, this reproduces eqn.~(11) of \citet{sph08}, except for
the autocorrelation term now intentionally omitted from \eqref{eqn:sigsdef2} (see Section~\ref{sec:perfect}) and 
the linear term now appended to \eqref{eqn:mdef} (see Section~\ref{sec:stepformv}).

We shall expand the (scale-dependent) mean squared error terms that reflect a measurement error, like $\langle|\bd\ep|^2\rangle$,
into a bias $\langle|\bd\ep|\rangle^2$, plus a covariance about the mean $\sigma^2[|\ep|]$.
For the sake of legibility, we do not likewise expand the mean squared error terms on instrument performance, such as $\langle|\ep|^2\rangle$, but the split is implicit.
For legibility, we also omit the notation showing functional dependence on scale and redshift, but note that all bias terms are functions of $(z_A,z_B)$ and all covariances are functions of $(\ell,z_A,z_B)$.
Indeed, since $\langle\rg\rangle$ scales with redshift, every term really will vary as a function of redshift. To second order in $\delta$, we find
\ba \label{eqn:sigsdef2_expand}
\add = \frac{1}{\pg^{~2}} \left\langle\frac{\rp^4}{\rg^4}\right\rangle 
\sigma^2[|\ep|] ~~~~~~~~~~~~~~~~~~~~~~~~~~~~~~ \nn
+ \frac{\langle|\ep|^2\rangle}{\pg^{~2}} \left\langle\frac{\rp^4}{\rg^4}\right\rangle \left(\frac{\langle\delta(\rp^2)\rangle^2}{\langle\rp^4\rangle} + \frac{\sigma^2[\rp^2]}{\rp^4} \right) \,\ea
where the spatially constant term in the first line disappears as a delta function at $\ell=1$ (or the fundamental mode of the survey) in Fourier space;
a similar cross term involving the (implicit) bias on $\ep$ is also zero in the second line.
Note that all ellipticities have two components that add in quadrature.
Ignoring a bias term in $\mult$ proportional to the square of one already present (therefore negligible if the bias is small), we also find
\be \label{eqn:mdef_expand}
\mult = 2\left\langle\frac{\rp^2}{\rg^2}\right\rangle\frac{\left\langle\delta(\rp^2)\right\rangle}{\left\langle\rp^2\right\rangle} +
\left\langle\frac{\rp^4}{\rg^4}\right\rangle \frac{\sigma^2[\rp^2]}{\left\langle\rp^4\right\rangle} \,.
\ee

We shall explore concrete values for the terms in equations~\eqref{eqn:sigsdef2_expand} and \eqref{eqn:mdef_expand} in Section~\ref{sec:amgeneral}.
For now, notice how the systematics are driven mainly by the size of the PSF --- to the fourth power, which is why cosmic shear measurements are generally easier from above the Earth's atmosphere.
However, $\delta(\rp^2)$ terms (proportional to only the second power) arise if the PSF is wavelength-dependent and measured from stars that are a different colour to galaxies \citep{ed09}.
This effect is worse for diffraction-limited space-based observations than ground-based imaging, where the PSF is determined primarily by atmospheric turbulence.
It would likely be spatially constant (and therefore disappear from $\add$ at least), except that chromatic aberration may exacerbate it on a characteristic scale related to the size of a telescope's field of view \citep{pla12}. It is anyway a function of redshift.

Equation \eqref{eqn:sigsdef2_expand} in particular shows that {\em overall performance is driven by the product of instrument stability and knowledge} about that instrument.
This quantifies the tradeoffs discussed by \citet{hardsoft}.
To obtain relaible cosmological measurements, we first need high-quality instrumentation to deliver a system PSF that is 
\begin{itemize}
\item small ($\rp$),
\item nearly circular (the bias component of $\langle|\ep|^2\rangle$) and 
\item stable (the variance component of $\langle|\ep|^2\rangle$; we have already assumed that its size is constant).
\end{itemize}
It is then equally important to 
\begin{itemize}
\item understand and accurately model that PSF.
\end{itemize}
The $\mean{\delta}$ terms reflect a calibration bias in the PSF model (\egg\ in its colour), and are likely to spatially constant.
The $\sigma^2[]$ terms reflect a lack of knowledge (\egg\ from sparse sampling of a spatially/temporally varying PSF pattern), and are likely to vary as a function of scale in such a way that they are largest around the mean distance between stars, the size of the telescope's field of view or (reflecting the intrinsic variation in the PSF pattern) turbulence cells in the atmosphere.

\subsection{Imperfect correction for detector effects} \label{sec:nl}

As well as convolution with a PSF (which in practice can include all optical and electronic effects that act linearly on pixel values),
astronomical images can also be degraded in more complicated ways.
This can include global detector nonlinearity, in which the number of counts in each pixel is a 
nonlinear function of the incident flux, or nonlocal effects
such as Charge Transfer Inefficiency in CCDs \citep{janesick} and inter-pixel capacitance or persistence in HgCdTe devices \citep{ipc1,ipc3,ipc4}. 

These operations cannot be treated mathematically as a convolution, so the correction procedure outlined in Section~\ref{sec:sph} does not apply.
We therefore introduce a new category of non-convolutive (NC) perturbations in galaxy size $\rc$ and ellipticity $\ec$.
The details of these may depend on the flux and size of the galaxy, but we take a generic approach (which can hold for small, faint galaxies) in which the observed quantities become
\be \label{eqn:rodef_rjm}
\ro\equiv\sqrt{\left(\rg^2+\rp^2\right)}+\rc
\ee
and
\be \label{eqn:eodef_rjm}
\eo\equiv\eg+\frac{\rp^2}{\rg^2+\rp^2}\left(\ep-\eg\right)+\ec~.
\ee
Note that we have not explicitly included non-convolution effects on stellar images from which the PSF is modelled.
The images of bright stars will also be degraded, and the budgets for $\delta\rp$ and $\bd\ep$ should allow for this.
However, many of the most serious nonlinear effects operate in the sense that the degradation of bright sources is much less than that of faint sources \citep{cti1,hoe11}.
In this case, the perturbations on galaxies $\rc$ and $\ec$ will dominate, in the budget for galaxy shape measurement rather than the (separable) budget for PSF modelling.
A weak lensing analysis then seeks to recover
\ba 
\eg = \left( 1+\frac{\rp^2}{\rg^2}\right) \left(\eo-\ec\right)-\frac{\rp^2}{\rg^2} \ep \label{eqn:egdef_int} ~~~ \\
 = \frac{\left(\eo-\ec \right)\left(\ro-\rc \right)^2-\ep\rp^2}{\left(\ro-\rc\right)^2-\rp^2}\,,\label{eqn:egdef_rjm}
\ea
where we have taken care on the second line to include only observable quantities on the right hand side.

In practice, any correction scheme will inevitably have inaccuracies $\delta\rc$ and $\bd\ec$, so only an imperfect estimation is possible of $\eg$.
Again we expand the shape observables as a first order Taylor series
\ba \label{eqn:eghatdef_rjm}
\widehat{\eg}\approx\eg + \frac{\partial\eg}{\partial(\rp^2)}\delta(\rp^2)
 + \frac{\partial\eg}{\partial\ep}\bd\ep ~+ \nn
   \frac{\partial\eg}{\partial(\rc)}\delta(\rc) 
 + \frac{\partial\eg}{\partial\ec}\bd\ec\,,~~~~
\ea
where the first two partial derivatives remain unchanged as (the second form of) equations~\eqref{eqn:partial1} and \eqref{eqn:partial2}, and
\ba 
\frac{\partial\eg}{\partial\rc} = \frac{2\rp^2\left(\ro-\rc\right)\left(\eo-\ec-\ep\right)}{\left[(\ro-\rc)^2-\rp^2)\right]^2} \nn
 = \frac{2\rp^2\left(\eg-\ep\right)}{\rg^2\sqrt{\left(\rg^2+\rp^2\right)}}~,~~~~~~~~~~~~~~~~~~~~~~~
\label{eqn:partial3}
\ea
\be
\frac{\partial\eg}{\partial\ec} = \frac{-(\ro-\rc)^2}{(\ro-\rc)^2-\rp^2}
 = -\frac{\rg^2+\rp^2}{\rg^2} ~. \label{eqn:partial4}
\ee
Thus
\ba \label{eqn:eghatdef3}
\widehat{\eg}\approx\eg\left\{1+\frac{\rp^2}{\rg^2}\left(\frac{\delta(\rp^2)}{\rp^2}+\frac{2\delta\rc}{\sqrt{\rg^2+\rp^2}}\right)\right\} \nn
     -~ \frac{\rp^2}{\rg^2}
       \Bigg\{\bd\ep + \frac{\rg^2+\rp^2}{\rp^2} \bd\ec + ~~~~~~~~\nn
       \left(\frac{\delta(\rp^2)}{\rp^2} + \frac{2\delta\rc}{\sqrt{\rg^2+\rp^2}}\right)\ep
       \Bigg\}.
\ea

This imperfect ellipticity measurement translates into additive cosmic shear systematics 
\ba 
\add = \frac{1}{\pg^{~2}} \left\langle\frac{\rp^4}{\rg^4}\right\rangle 
\sigma^2[|\ep|] ~~~~~~~~~~~~~~~~~~~~~~~~~~~~~~~ \nn
+ \frac{1}{\pg^{~2}} \left\langle\left(1+\frac{\rp^2}{\rg^2}\right)^2~\right\rangle 
\sigma^2[|\ec|] ~~~~~~~~~~~~~~~~~~~ \nn
+ \frac{\langle|\ep|^2\rangle}{\pg^{~2}} \left\langle\frac{\rp^4}{\rg^4}\right\rangle \left(\frac{\langle\delta(\rp^2)\rangle^2}{\langle\rp^4\rangle} + \frac{\sigma^2[\rp^2]}{\rp^4} \right) \, \nn
+ \, 4 \frac{\langle|\ep|^2\rangle}{\pg^{~2}} \left\langle\frac{\rp^4}{\rg^4}\right\rangle \left(\frac{\langle\delta\rc\rangle^2}{\langle\rc^2\rangle} + \frac{\sigma^2[\rc]}{\rc^2} \right) ~~~
\ea
Mixing thus emerges between corrections for convolution and non-convolution effects.
In the second term for example, imperfections $\bd\ec$ in the correction for
detector effects are enhanced during subsequent deconvolution.

The multiplicative cosmic shear systematics
\ba 
\mult = 
2\left\langle\frac{\rp^2}{\rg^2}\right\rangle \Bigg( \frac{\left\langle\delta(\rp^2)\right\rangle}{\left\langle\rp^2\right\rangle} + 2\frac{\left\langle\delta(\rc)\right\rangle}{\left\langle\ro\right\rangle} \Bigg) \nn
+
\left\langle\frac{\rp^4}{\rg^4}\right\rangle \Bigg( \frac{\sigma^2[\rp^2]}{\left\langle\rp^4\right\rangle} + 4\frac{\sigma^2[\rc]}{\left\langle\ro^2\right\rangle} \Bigg)~~~
\ea
plus bias terms proportional to the square of those already present (therefore negligible if the bias is small).
The second term $\mean{\delta\rc}$ 
reflects overall uncertainty in the model of non-convolution effects, such as the density and characteristic release time of charge traps in CCDs.
These quantities may be stable over long periods of time, but the error may vary as a function of object flux (hence redshift) if, in this case, the CCD well-filling model is inaccurate.
The fourth term $\sigma^2[\delta\rc]$ reflects unaccounted variation of an effect at different positions within a detector. 
Depending on survey tiling strategies, NC terms are likely to be largest on physical scales corresponding to linear multiples of the chip size \citep[see][]{cropper12}.

\subsection{Imperfect shape measurement methods} \label{sec:step}

In the previous sections we examined the impact of errors in the measurements of the 
PSF and detector effects, but we implicitly assumed that the observed galaxy moments
are unbiased. In practice, the unweighted size $\ro$ and shape $\eo$ of a faint galaxy may be subject to 
errors $\delta\ro$, $\bd\eo$ for a whole variety of reasons including mis-centering, background gradients/structure, 
pixellisation, and simply noise. We therefore need to consider also the impact of imperfections in the
measurements of the galaxies. This leads to new contributions to the observed ellipticity
\ba \label{eqn:eghatdef_smm}
\bmath{\widehat\gamma}\equiv(\widehat{\pg})^{-1}~\widehat{\eg} ~~~~~~~~~~~~~~~~~~~~~~~~~~~~~~~~~~~~~~~~~~~~~\, \\
\approx\frac{\eg}{\pg} + \frac{1}{\pg}\frac{\partial\eg}{\partial(\rp^2)}\delta(\rp^2)
 + \frac{1}{\pg}\frac{\partial\eg}{\partial\ep}\delta\ep \nn
 + \frac{1}{\pg}\frac{\partial\eg}{\partial(\rc)}\delta(\rc) 
 + \frac{1}{\pg}\frac{\partial\eg}{\partial\ec}\delta\ec ~~~~~~~~~ \nn
 + \frac{1}{\pg}\frac{\partial\eg}{\partial(\ro^2)}\delta(\ro^2) 
 + \frac{1}{\pg}\frac{\partial\eg}{\partial\eo}\delta\eo - \frac{\delta\pg}{\pg^{~2}}\eg\,.\hspace{-7mm}
\ea
The new partial derivatives of \eqref{eqn:egdef_rjm} are
\be \label{eqn:partial5}
\frac{\partial\eg}{\partial(\ro^2)} = -\frac{\rp^2}{\rg^2}\frac{(\eg-\ep)}{\ro(\ro-\rc)}~,
\ee
\be \label{eqn:partial6}
\frac{\partial\eg}{\partial\eo} = \frac{\rg^2+\rp^2}{\rg^2} ~.
\ee
Alternatively, note that $\partial\eg/\partial\ro=-\partial\eg/\partial\rc$.
Including observational error, we thus find the shear measurement $\widehat{\bmath\gamma}$ has biases given by STEP parameters
(eqn.~\ref{eqn:stepgamma}) 
\ba 
\bc = \frac{1}{\pg}\frac{\rp^2}{\rg^2}
       \Bigg\{\frac{\rg^2+\rp^2}{\rp^2}\left(\bd\eo-\bd\ec\right) - \bd\ep ~~ \\
    - \left(\frac{\delta(\rp^2)}{\rp^2} +
      \frac{2\,\delta\rc}{\ro-\rc} +
  \frac{\delta(\ro^2)}{\ro(\ro-\rc)}\right)\ep
      \Bigg\}\hspace{-8mm} \nonumber
\ea
\ba 
m=\frac{\rp^2}{\rg^2}\Bigg\{\frac{\delta(\rp^2)}{\rp^2} 
    + \frac{2\,\delta\rc}{\ro-\rc} - \frac{\delta(\ro^2)}{\ro(\ro-\rc)} \Bigg\} \hspace{-70mm} \nn -\frac{\delta\pg}{\pg}.
\ea

We have so far considered only shape measurement using unweighted moments.
This approach greatly simplifies the calculations, but potentially limits the applicability to real data. 
This is because the presence of  any noise in an image formally leads to infinite noise in the measurements of unweighted moments. 
It may be feasible to measure (close to) unweighted moments in the special cases of very bright stars, or of repeated detector effects, by stacking data to suppress the noise.

It is never possible in practice to measure directly the unweighted moments of faint galaxies, and one has to use weighted moments instead. 
The optimal weight function to use, is the one that maximizes the signal-to-noise ratio, which in turn implies that the weight function closely resembles the galaxy profile. This is naturally done by methods that fit  parametric shape models to the data \citep[\egg][]{im2shape,lensfit1,lensfit2}. 
Moment based methods (\egg\ \citealt{ksb,rrg}, hereafter KSB and RRG) instead construct sizes $\wrg$ and ellipticities $\weg$ from quadrupole moments weighted by a radial Gaussian function, the size of which is matched to the object.
There are no simple expressions that relate $\rg$ and $\eg$ in terms of observed weighted moments, equivalent to the unweighted versions \eqref{eqn:rodef_rjm} and \eqref{eqn:eodef_rjm}.
Derivations using weighted moments are complicated and involve mixing of higher order moments \citep[KSB; RRG;][]{k2k,shapelets1,deimos}.
For any individual galaxy however, it is possible\footnote{For example, the shear estimator in KSB (in the absence of non-convolutive effects) is
\be
\widehat{\gamma}_{\rm w} = \left(P^\gamma_{\rm KSB} \right)^{-1} \left[\weo - P^{\rm sm}\left(P^{\rm sm}_{\rm PSF} \right)^{-1}\, \wep \right], \nonumber
\label{eqn:ksb}
\ee
where
\be
P^\gamma_{\rm KSB} = P^{\rm sh} - P^{\rm sm} \left(P^{\rm sm}_{\rm PSF} \right)^{-1} P^{\rm sh}_{\rm PSF}. \nonumber
\label{eqn:ksbPgamma}
\ee
The interpretation of such quantities is method-specific.
If $\widehat{\bmath{\gamma}}_{\rm w} \sim (\pg)^{-1}(\peo^\prime)^{-1}\weo$,
the middle factor can be interpreted as part of either the polarizability \citep[\egg][use higher order moments to construct 
$(\pg\peo^\prime)^{-1}$]{ksb,shapelets4}, or as part of  the ellipticity \citep[\egg][use higher order moments to convert 
weighted ellipticities to unweighted ellipticities $(\peo^\prime)^{-1}\weg$]{rrg,k2k}.
} to define without loss of generality various $P^\prime$ quantities to form a shear estimator from weighted moments
\ba
\widehat{\gamma}_{\rm w} \equiv \frac{1}{\pg}\left(1+\frac{1}{\pr^\prime}\frac{\wrp^2}{\wrg^2}\right) \left(\frac{\weo}{\peo^\prime}-\frac{\wec}{\pec^\prime}\right) ~~~~~~~~ \nn
- \frac{1}{\pg}\frac{1}{\pr^\prime}\frac{\wrp^2}{\wrg^2}\left(\frac{\wep}{\pep^\prime}\right), \label{eqn:egweighted}
\ea
where we have intentionally arranged terms to resemble equation~\eqref{eqn:egdef_int}.
For individual galaxies, especially those with complex intrinsic shapes, it can be that $\widehat{\gamma}_{\rm w}\ne \widehat{\gamma}$, as long as averaged over a large population of galaxies, $\langle\widehat{\gamma}_{\rm w}\rangle=\langle\widehat{\gamma}\rangle$.

The $P^\prime$ quantities fulfil two roles, and can even be expressed as the product of discrete quantities 
\be
P^\prime_x=W_xP_x.
\ee
Both components of $P^\prime_x$ are tensors, but they are nearly diagonal, so for simplicity we shall treat them as scalars.
The first component $W_x$ compensates for the weight function's changes to moments, \egg
\ba
\frac{\wrp^2}{\wrg^2}&\equiv&{W_R}\,\frac{\rp^2}{\rg^2} \label{eqn:wrg} \\
\wep&\equiv&\pwep~\ep \label{eqn:wep} 
\ea
\etc. Numerical values of this component depend upon the shape measurement method, but for small galaxies $W_R\sim1$ as it governs a ratio of similar quantities and $W_\ellipticity\sim1/2$ (for all the ellipticities).
The second component $P_x$ encodes the way in which the effective PSF is altered by the weight function, and its numerical values depend upon the PSF properties.
For a Gaussian PSF, all $P$ values are exactly equal to $1$.
This approximately holds for a smooth (\egg\ ground-based) PSF or a small PSF (or a large galaxy).
For an Airy PSF, the outer diffraction wings are damped by the weight function\footnote{Consider the pathological example of a PSF consisting of a core plus a ring at large radius. The ring lowers the perceived flux of a galaxy, but has no effect on its size or shape as determined from weighted moments.}, leading to large differences between weighted and unweighted quantities.
For large galaxies, the weight function will be extended and the suppression is small. 
For small galaxies, size estimates are most affected, and we find $\pr\sim2$: the net effect of the weight function is equivalent to reducing the PSF size.
Ellipticities are less affected, with $\pw\sim1$ in any observing regime.
This depends weakly on the intrinsic ellipticity and size but, since we shall generally consider limiting cases of small/faint galaxies, we shall henceforth treat these factors as constants.

We now re-evaluate the additive and multiplicative biases, accounting for the use of weighted moments.
This could involve replacing all mentions of observable sizes and shapes by their weighted equivalents. 
However, for comparison with our earlier results, and to eventually express engineering requirements on instrumentation, it is more convenient to continue to use unweighted quantities. 
Substituting equations~\eqref{eqn:wrg} and \eqref{eqn:wep} into \eqref{eqn:egweighted}, we find
\ba
\widehat{\gamma}_{\rm w} \equiv \frac{1}{\pg}\left(1+\frac{1}{\pr}\frac{\rp^2}{\rg^2}\right) \left(\frac{\eo}{\peo}-\frac{\ec}{\pec}\right) ~~~~~~~~~~~ \nn
- \frac{1}{\pg}\frac{1}{\pr}\frac{\rp^2}{\rg^2}\left(\frac{\ep}{\pep}\right), \label{eqn:egweighted2}
\ea
This expression clearly demonstrates how weighted moments can naturally suppress bias. 
However, this advantage comes at a price.
The evaluation of the $P$ factors requires knowledge of higher order shape moments, which can be well known for bright stars but are especially noisy for faint galaxies.
The absolute values of $\pep$, $\pec$ and $\peo$ adjust the balance between different contributions to the bias, but errors in those quantities are functionally identical to errors in $\ep$, $\ec$ and $\eo$, which we have already considered.
Observational errors in $\pr$ propagate into a new source of bias, via 
\ba 
\bmath{\widehat{\gamma}}_{\rm w}\approx \bmath{\gamma}_{\rm w} + \frac{\partial\,\bmath{\gamma}_{\rm w}}{\partial(\rp^2)}\delta(\rp^2)
 + \frac{\partial\,\bmath{\gamma}_{\rm w}}{\partial\ep}\delta\ep ~~~~ \nn
 + \frac{\partial\,\bmath{\gamma}_{\rm w}}{\partial(\rc)}\delta(\rc) 
 + \frac{\partial\,\bmath{\gamma}_{\rm w}}{\partial\ec}\delta\ec ~~~~~ \nn
 + \frac{\partial\,\bmath{\gamma}_{\rm w}}{\partial(\ro^2)}\delta(\ro^2) 
 + \frac{\partial\,\bmath{\gamma}_{\rm w}}{\partial\eo}\delta\eo \nn
 + \frac{\partial\,\bmath{\gamma}_{\rm w}}{\partial\pr}\delta(\pr)
 - \frac{\delta\pg}{\pg}\bmath{\gamma}_{\rm w}\,, ~~~
\ea
where the derivatives of $\bmath{\gamma}_{\rm w}$ gain prefactors of $1/\pr$ or $1/\pr P_\ellipticity$ compared to those of $\bmath{\gamma}$ and
\be
\frac{\partial\,\bmath{\gamma}_{\rm w}}{\partial\pr} = -\frac{1}{\pr}\left( \frac{\rg^2}{\pr\rg^2+\rp^2} \right) \left( \bmath{\gamma}_{\rm w} +\frac{\ep}{\pg\pep} \right) .
\ee

If the size of the PSF depends upon wavelength, this term introduces a sensitivity to spatial variations in the colour of a galaxy \citep[whereby the PSF is different in the bulge and the disc:][Semboloni et al.\ in prep.]{voigt11}.
This is because multiple galaxy profiles result in galaxies with identical observed moments, so the estimate for $\pr$ becomes biased.
Similar biases in $\pr$ also arise in parametric fitting methods if the model does not reflect galaxies' true morphological characteristics \citep{voigt10}, suffers from aliasing \citep{ber10}, or is nonlinear \citep{ref12}.
In this paper we do not distinguish between these individual origins, but consider all such effects part of a general method bias.

We conclude that a shear estimator $\bmath{\widehat{\gamma}}_{\rm w}$ constructed  from weighted moments has STEP biases
\ba \label{eqn:stepclast}
\bc = 
     \frac{1}{\pg\pr} \frac{\rp^2}{\rg^2} \Bigg\{
      \left(\frac{\pr\rg^2+\rp^2}{\rp^2}\right)\left(\frac{\bd\eo}{\peo}-\frac{\bd\ec}{\pec}\right) \hspace{-3mm} \nn 
    - \frac{\bd\ep}{\pep} 
    - \frac{\ep}{\pep} \Bigg(\frac{\delta(\rp^2)}{\rp^2} + 
      \frac{2\,\delta\rc}{\ro-\rc} + ~~~\nn
  \frac{\delta(\ro^2)}{\ro(\ro-\rc)} 
    +  \frac{\pr\rg^4}{\rp^2(\pr\rg^2+\rp^2)}\frac{\delta\pr}{\pr} \Bigg)  \Bigg\} 
\ea
\be \label{eqn:stepmlast}
m=\frac{1}{\pr}\frac{\rp^2}{\rg^2}\left(\frac{\delta(\rp^2)}{\rp^2} 
    + \frac{2\,\delta\rc}{\ro-\rc} + \mu\right),
\ee
where
\ba
\mu\equiv - \frac{\delta(\ro^2)}{\ro(\ro-\rc)} ~~~~~~~~~~~~~~~~~~~~~~~~~~~~~~~~~~ \nn
  - \pr\frac{\rg^2}{\rp^2} \left\{ \frac{\delta\pg}{\pg} +
    \left(\frac{\rg^2}{\pr\rg^2+\rp^2}\right)\frac{\delta\pr}{\pr} \right\}
\ea
is the component of bias due to the galaxy shape measurement method.
The STEP parameter ${\bmath{q}}$, which flags an (incorrect) quadratic response to shear, could be produced by measurement errors that depend on intrinsic ellipticity such as $\bd\eo(\eg)$. Averaged over a galaxy population, these are functionally identical to errors $\delta\pg$.

Observational errors are likely isotropic \ie\ $\langle\bd\eo\rangle=0$ and spatially constant \ie\ in the absence of galaxy-galaxy auto\-correlations $\sigma^2[\ro^2]=\sigma^2[|\eo|]=\sigma^2[\pg]=\sigma^2[\pr]$=0. This general case thus has additive cosmic shear systematics
\ba \label{eqn:sigsdef2_exp}
\add = \frac{1}{\pr^{~2}\pg^{~2}} \left\langle\frac{\rp^4}{\rg^4}\right\rangle
\frac{\sigma^2[|\ep|]}{\pep^{~2}} ~~~~~~~~~~~~~~~~~~~~~~~~~~~~~~~~~ \hspace{-10mm} \nn
+ \frac{1}{\pr^{~2}\pg^{~2}} \Bigg\langle\Bigg(\pr^{~2}+\frac{\rp^2}{\rg^2} \Bigg)^2\, \Bigg\rangle\,
\frac{\sigma^2[|\ec|]}{\pec^{~2}} ~~~~~~~~~~ \nn
+ \frac{ \langle|\ep|^2\rangle}{\pr^{2}\pg^{2}\pep^{~2}} \left\langle\frac{\rp^4}{\rg^4}\right\rangle \left( \frac{\langle\delta(\rp^2)\rangle^2}{\langle\rp^4\rangle} + \frac{\sigma^2[\rp^2]}{\rp^4} \right)  \hspace{-5mm} \nn
+ \,  \frac{4\,\langle|\ep|^2\rangle}{\pr^{2}\pg^{2}\pep^{~2}} \left\langle\frac{\rp^4}{\rg^4}\right\rangle \left( \frac{\langle\delta\rc\rangle^2}{\langle\rc^2\rangle} + \frac{\sigma^2[\rc]}{\rc^2} \right) ~~~~ \hspace{-5mm} \nn
+ \frac{\langle|\ep|^2\rangle}{\pr^{2}\pg^{2}\pep^{~2}}\left\langle\frac{\rp^4}{\rg^4}\right\rangle \alpha^2
~~~~~~~~~~~~~~~~~~~~~~~~~~~~~~~~~ \hspace{-5mm} 
\ea
where
\be
\alpha^2\equiv
   \frac{\langle\delta(\ro^2)\rangle^2}{\langle\ro^4\rangle} 
+ \left\langle\frac{\rg^4}{\rp^4}\right\rangle \Bigg\langle \Bigg( \frac{\pr\rg^2}{\pr\rg^2+\rp^2} \Bigg)^2\, \Bigg\rangle\,\frac{\langle\delta\pr\rangle^2}{\langle\pr^2\rangle}\,. ~
\ee
The first term in $\alpha^2$ could arise due to pixellisation effects, but this will be zero for resolved imaging and deviations could be measured only by changing the plate scale in a camera.
Note that if autocorrelations of galaxy shapes with themselves are included in the correlation function analysis (see Section~\ref{sec:perfect}), the additive cosmic shear systematics gain an extra white noise term $\sigma^2_{\bmath{\gamma}}(z_A,z_B)$ as in equation~\eqref{eqn:siggamstep}. 
The multiplicative cosmic shear systematics become
\ba \label{eqn:mdef_exp}
\mult = 
\frac{2}{\pr}\left\langle\frac{\rp^2}{\rg^2}\right\rangle \Bigg( \frac{\left\langle\delta(\rp^2)\right\rangle}{\left\langle\rp^2\right\rangle} + 2\frac{\left\langle\delta(\rc)\right\rangle}{\left\langle\ro\right\rangle} +\langle\mu\rangle \Bigg) \nn
+\frac{1}{\pr^{~2}}\left\langle\frac{\rp^4}{\rg^4}\right\rangle \Bigg( \frac{\sigma^2[\rp^2]}{\left\langle\rp^4\right\rangle} + 4\frac{\sigma^2[\rc]}{\left\langle\ro^2\right\rangle} \Bigg). ~~~~~~~~~
\ea

\begin{figure*}
 \includegraphics[width = 17cm]{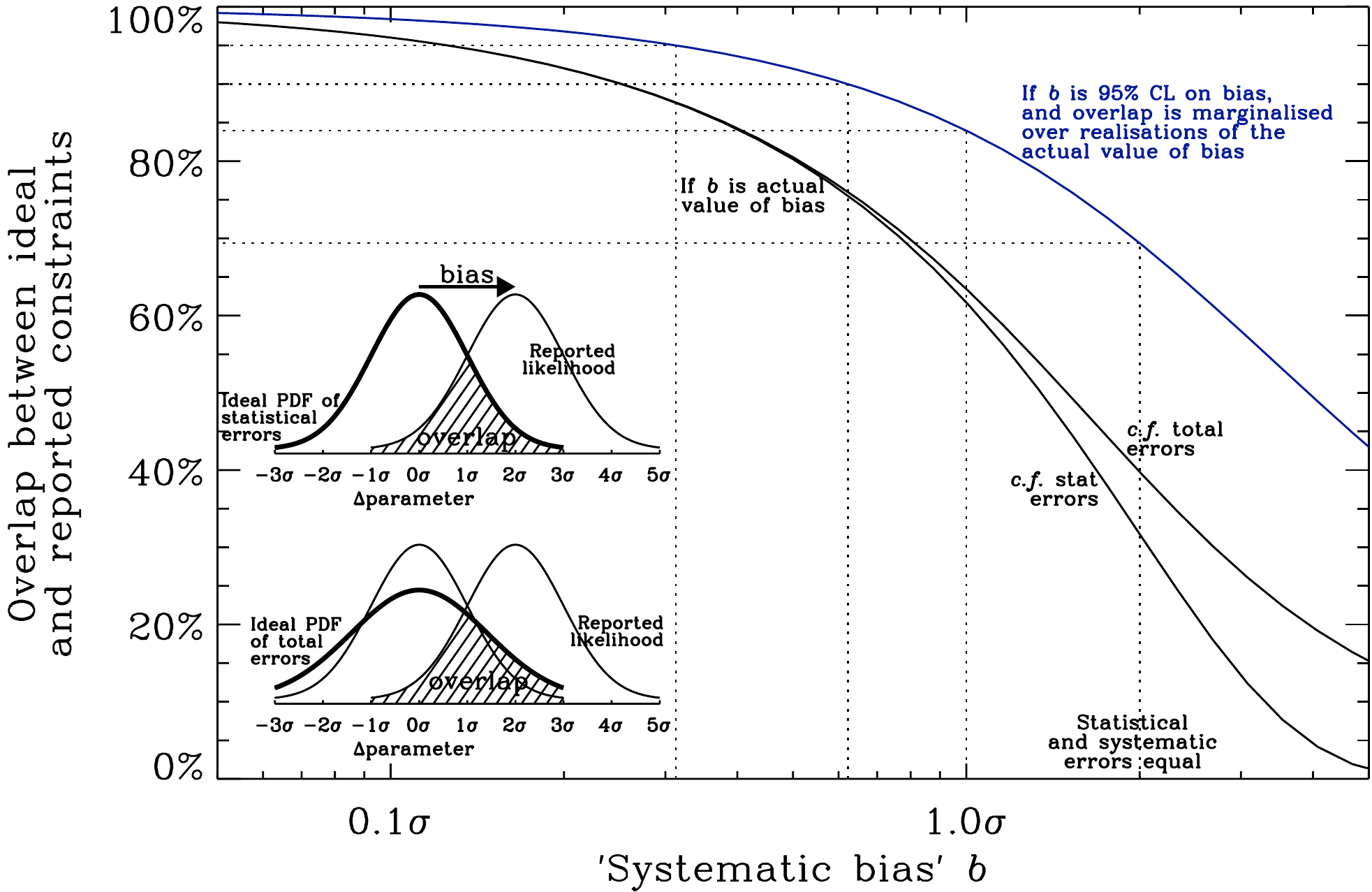}
\caption{\label{fig:overlaps}
The effect of a systematic bias in an experiment, as a function of statistical error $\sigma$, assuming all likelihood distributions are Gaussian.
The $y$ axis is the chance that a reported experimental result is drawn from the reported likelihood distribution, centered around the true value.
The two lower (black) curves show this chance in the presence of some exact systematic bias $b$, for the reported statistical errors (lower) or total errors (upper).
The top (blue) curve shows this chance if $b$ is instead a $95$\% confidence limit on the (unknown) true bias.
In this case, the true bias {\em could} be zero, so the overlap with the ideal likelihood distribution is always greater.
Using this latter definition, we require $b<0.31\sigma$ for a $95$\% overlap with the ideal PDF.}
\end{figure*}

\section{Requirements to meet future science goals} \label{sec:cosmo}

\subsection{How much systematic bias is tolerable?}
\label{sec:howmuch}

Total experimental error from any measurement is always a combination of systematic and statistical errors. 
Systematic bias (\ie\ the deviation of a measured value from the truth) can be reduced by \egg\ stabilising a telescope or raising it above the Earth's atmosphere. 
Statistical error (\ie\ the confidence interval allocated to a measured value) is limited by the finite number of measurements within a survey, and can be reduced by \egg\ increasing survey volume. 
The diagrams inset within Figure~\ref{fig:overlaps} illustrate how an unrecognised systematic bias shifts measurements, which are drawn from a statistical likelihood distribution around the offset value. 

Classical astronomical survey design optimises an observation that is limited by systematic biases inherent to a technique or its interpretation.
This limitation drives surveys wider, deeper or to higher resolution, until their statistical error becomes smaller than the systematic bias.
However, several surveys planned for the next decade have scientific goals that require them to image the entire sky outside the plane of the Milky Way.
Further increasing survey area is impossible, and increasing survey depth can be prohibitively expensive: especially for space-based surveys, where mission cost jumps in step functions with mirror size (bigger launch vehicles become necessary) or survey duration (additional redundancy of components). 
For these surveys, the statistical error is fixed and the classical trade-off is inverted;
the key question becomes {\em how much systematic bias is tolerable?} 
We shall answer this quantitatively by considering the probability with which an experiment's reported measurement of a particular parameter could have been obtained by an unbiased experiment. 
This is the overlap integral between the likelihood distribution reported by an experiment, and the likelihood distribution that would have been reported by an unbiased experiment (\ie\ the same distribution, re-centered around the parameter's true value)\footnote{This is a frequentist argument based on $p$-value like statistics; a Bayesian methodology, in which evidence ratios for a target model with and without systematics could also be considered.
We shall also only consider the 1 dimensional bias on a single parameter at a time \citep[\cf][]{dod06,sha09,sha10}.}.

Throughout this section, we shall consider statistical errors described by a Gaussian of width $\sigma$. 
There are two ways in which a systematic bias can be described. 
Following frequent use in the literature, and as illustrated in the upper inset panel of figure~\ref{fig:overlaps}, we shall first consider an experiment with an exact amount of bias, $b$.
The lowest curve in Figure~\ref{fig:overlaps} shows the probability that a reported measurement could have been sampled from the unbiased (re-centered) likelihood. 
This is simply the (cross-hatched) overlap integral under two Gaussians with variance $\sigma_1=\sigma_2=\sigma$ and mean $\mu_1=0$, $\mu_2=b$
\ba \label{eqn:overlap1}
p_{\rm overlap}^{\rm stat}(b)=
\int_{-\infty}^\infty
\mathrm{min}\left\{
\frac{{\rm e}^\frac{-x^2}{2\sigma^2}}{\sqrt{2\pi\sigma^2}}~,
\frac{{\rm e}^{-\frac{(x-b)^2}{2\sigma^2}} }{\sqrt{2\pi\sigma^2}}
\right\}{\rm d}x~~~~~~~~~~\\
=\frac{1}{\sqrt{2\pi\sigma^2}} \left(
\int_{-\infty}^{\frac{|b|}{2}}{\rm e}^{\frac{-(x-|b|)^2}{2\sigma^2}}{\rm d}x +
\int_{\frac{|b|}{2}}^{\infty}{\rm e}^{-\frac{x^2}{2\sigma^2}}{\rm d}x \right) \\
=1-{\rm erf}\left(\frac{1}{2\sqrt{2}}\frac{|b|}{\sigma}\right). ~~~~~~~~~~~~~~~~~~~~~~~~~~~~~~~~~~
\ea
For the overlap to be at least $95$\% ($90$\%), the absolute value of bias $|b|$ must be less than $0.13\sigma$ ($0.25\sigma$).
If bias is allowed to be as large as the $1\sigma$ statistical error, the overlap integral is only $62$\%, which is undesirable.
One effect slightly improves this: as illustrated in the lower inset diagram, reported error bars will be enlarged to account for an estimate of the systematic bias.
The middle curve in Figure~\ref{fig:overlaps} shows what happens if the achieved level of bias were treated as a $95$\% confidence limit on a Gaussian systematic error budget, \ie\ $\sigma_b=b/2$.
In this case, the overlap integral becomes
\be \label{eqn:overlap2}
p_{\rm overlap}^{\rm total}(b)=\int_{-\infty}^\infty
\mathrm{min}\left\{
\frac{{\rm e}^\frac{-x^2}{2(\sigma^2+\sigma_b^2)}}{\sqrt{2\pi(\sigma^2+\sigma_b^2)}}~,
\frac{{\rm e}^{-\frac{(x-b)^2}{2\sigma^2}} }{\sqrt{2\pi\sigma^2}}
\right\}{\rm d}x
\ee
although this does not significantly affect $p$.

However, any systematic bias that is known exactly would already have been subtracted from a measurement!
We shall now re-interpret $b$ as the $95$\% confidence limit on the absolute value of an {\em unknown} bias.
A Gaussian distribution of possible biases with mean zero and width $\sigma_b=b/2$ sometimes creates small or even zero bias, so the overlap of reported and ideal measurements is greater.
Marginalising over this distribution, the top curve in Figure~\ref{fig:overlaps} shows
\be
p_{\rm overlap}^{\rm marginalised}(b)=\frac{1}{\sqrt{2\pi\sigma_b^2}}\int_{-\infty}^\infty e^{\frac{-b^{\prime 2}}{2\sigma_b^2}}~p_{\rm overlap}^{\rm total}(b^\prime)~ {\rm d}b^\prime.
\ee
Achieving a $95$\% ($90$\%) probability that a reported result could have been drawn from the likelihood distribution re-centered on the true value now requires $|b|<0.31\sigma$ ($0.62\sigma$). 
Only $69$\% overlap arises if the systematic and statistical error budgets are equal ($\sigma_b=\sigma$). 
We shall henceforth require uncertain biases to have a $95$\% confidence limit that is less than $31$\% of the $1\sigma$ expected statistical error.

\subsection{Propagation of shear measurement errors to biases on cosmological parameters} \label{sec:sims}

We now propagate hypothetical shear measurement errors $\add(\ell,z_A,z_B)$ and $\mult(\ell,z_A,z_B)$ 
from Section~\ref{sec:problems} into biases on derived cosmological parameters, via the Fisher matrix bias formalism (\citealt{fisher}; see also \citealt{ar07}, \citealt{icosmo2}).
In particular, we concentrate on measurements of the dark energy equation of state parameter $w$ or its derivative $w_a$ \citep{che01,lin03}, and marginalise over other parameters. 
By requiring that the $95$\% confidence limit on bias is less than $31$\% of the 
statistical errors afforded by Poisson noise in a finite survey volume (see Section~\ref{sec:howmuch}), we obtain requirements
on the accuracy with which the PSF must be modelled, 
detector effects must be corrected, and galaxy shapes must be measured.
This is more stringent than the work of \citet{ar08}, who required bias less than $100\%$ of statistical error.

We assume a baseline $15,000$ square degree cosmic shear survey 
resolving $30$ galaxies per square arcminute with median redshift of $1.0$ and split into $10$ tomographic redshift bins.
This matches the configuration of the proposed Euclid mission \citep{euclidredbook}, and is likely to be similar to any proposed Stage~IV survey: for example, LSST proposes to survey $18,000$ square degrees with an effective density of $40$ galaxies per square arcminute \citep{jeety11,bra12}.
\citet{das12} describe the effect of perturbing the parameters of the baseline survey in a similar analysis. 

We use the {\tt iCosmo} Fisher matrix software \citep{icosmo1,icosmo2} to calculate the concordance $\Lambda$CDM cosmic shear power spectrum $C(\ell,z_A,z_B)$ in a top-hat basis set (200 bins) spanning scales $10<\ell<5000$ and every pair of redshift bins.
Henceforth, $\ell$, $z_A$ and $z_B$ refer to the median values of the population of galaxies within these bins.
We assume the Limber approximation, and neglect any power spectrum due to intrinsic alignments.
Using only weak lensing measurements, such an experiment can measure $w$ with a 1$\sigma$, 1-parameter statistical error of $0.065$, and $w_a$ with a statistical error of $0.41$.

\subsection{Constant additive and multiplicative shear measurement bias} \label{sec:amconst}

To first explore the consequences of the simplest possible systematic errors, we first impose upon each measurement of $C(\ell)$
a constant additive shear measurement bias $\add$ (or $\sigc $) and
a constant multiplicative shear measurement bias $\mult$ (or $m$).
This simultaneity of multiplicative and additive biases has not been explored before, with previous studies in the literature considering the imposition of only one type of systematic at a time.
Note that although $\sigma_{\bc}^2$ is positive by definition, and $m$ is almost always negative in practice \citep[\egg][]{great08}, we explore positive and negative values in both cases because {\em if their values are known, they would be removed from data} (or added to models). 
The {\em only} important parameter is the residual after this process, \ie\ the accuracy to which  $\add$ and $\mult$ are known. By definition, this residual is equally likely to be either positive or negative.

\begin{figure}
\includegraphics[width = 8.5cm]{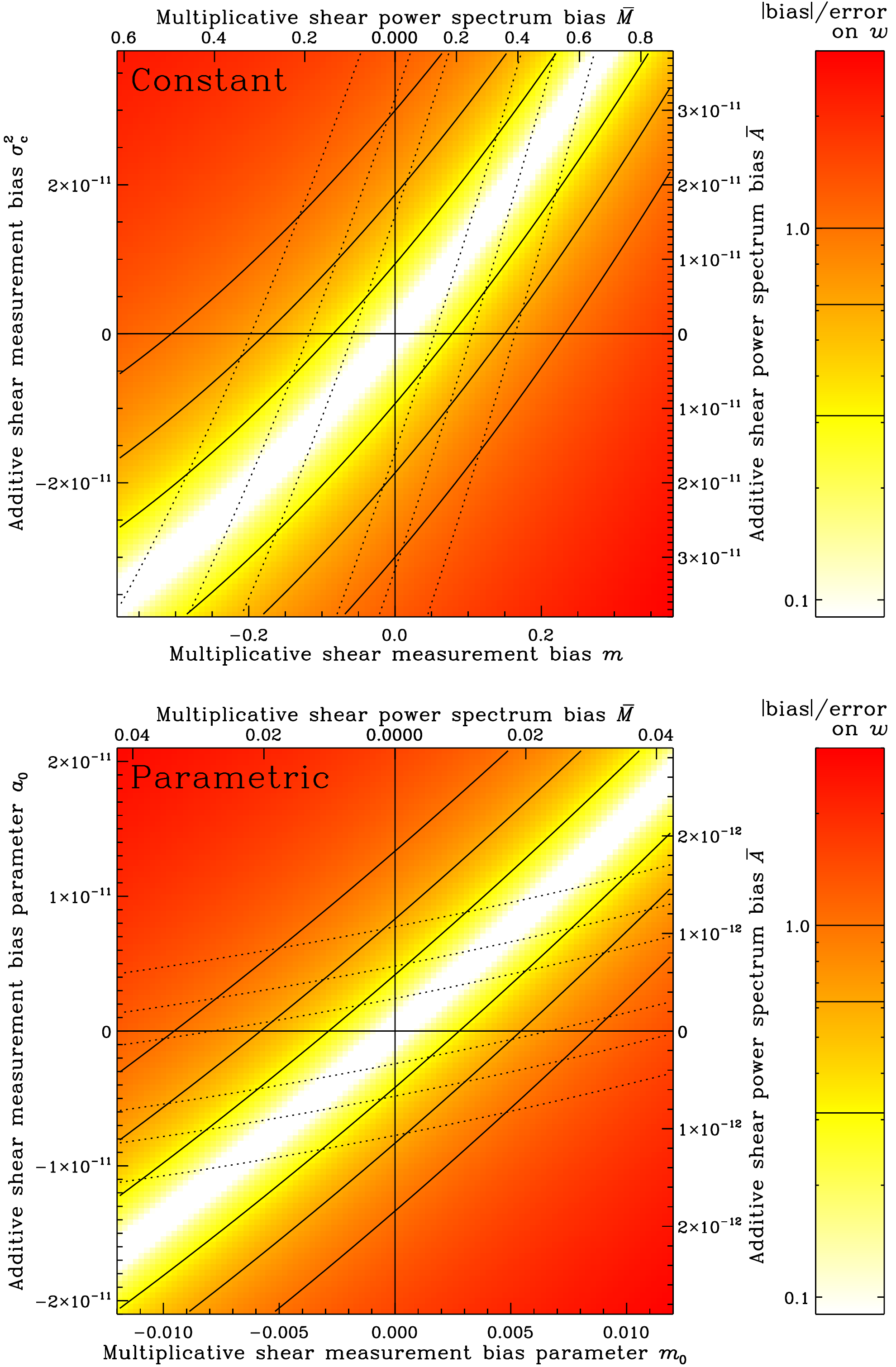}
\caption{\label{fig:mcsys}
The (absolute value of) bias on measurements of the dark energy equation of state parameters $w$ (colour and solid contours) and $w_a$ (dotted contours) 
from weak lensing surveys with multiplicative and additive shear measurement systematics.
Bias is shown as a multiple of the expected statistical error $\sigma$,
and contours are drawn at the same values as in Figure~\ref{fig:overlaps}.
{\bf Top panel:} constant systematics $m$ and $\sigma_{\bc}^2$.
{\bf Bottom panel:} one realisation of variable systematics $\mult(z)$ and $\add(\ell)$, as described in the text
(note the change of scale).}
\end{figure}

We find that there is a degeneracy between the two types of bias,
in terms of the way they influence constraints on the dark energy equation of state parameter $w$ (Figure~\ref{fig:mcsys} top panel).
Indeed, if $\add$ and $\mult$ have the same sign, they can cancel each other out to produce no net bias on $w$. 
However, the tuning of this cancellation is specific to the parameter being measured: the degeneracy is completely different for measurements of $w_a$, $\Omega_m$ or $\sigma_8$.

Given our first order expansion, it is not surprising that the surface of figure~\ref{fig:mcsys} is approximately fitted by a plane 
$b/\sigma\approx-0.093-3.9m+3.3\times10^{10}\sigc $.
Thus, if the signs of $\add$ and $\mult$ are not known {\it a priori}, guaranteeing $|b|<0.31\sigma$ requires
\be
|m|+8.6\times10^9|\sigc |\simlt0.10\,.
\ee
Whilst surprisingly large constant $m$ can be acceptable for measurements of $w$ (since $\mult C(\ell)$ does then not resemble $\partial C(\ell)/\partial w$), we again note that this is not true for measurements of other cosmological parameters.
Most importantly, we note the necessity for joint requirements on $\add$ and $\mult$.
Whenever requirements are placed on $\add$ when assuming $\mult\equiv0$ or vice versa, one degenerate error budget is being spent twice. 
The two requirements should be halved and, since the bias surface is well-fit by a plane, the two requirements can be linearly traded against each other.
This degeneracy has not been taken into account by earlier work.

\subsection{Simple forms of additive and multiplicative shear measurement bias} \label{sec:amsimple}

As discussed in Section~\ref{sec:problems}, systematics often affect some physical scales more than others, and it is typically more difficult to measure the shapes of distant (small, faint) galaxies than nearby (big, bright) ones.
One feasibly more realistic functional form for non-constant additive systematics is
\be
\add(\ell)=a_0 \left(1+\frac{\ell}{\ell_0}\right)^{\beta_2-\beta_1} \left(\frac{\ell}{\ell_0}\right)^{\beta_1} , \label{eqn:Aparametric}
\ee
where $\ell_0=1000$, $\beta_1=-1.5$, $\beta_2=-3$ \citep[eqn.\ 29 of][]{hardsoft}.
A feasible functional form for multiplicative systematics is
\be
\mult(z_A,z_B)=m(z_A)+m(z_B)+m(z_A)\times m(z_B), \label{eqn:Mfromm}
\ee
where
\be
m(z)=m_0 \frac{2}{\pi}(1+z)^{\beta_m}~\mathrm{tan}^{-1}\big(\alpha_m(z-z_T)\big)  \label{eqn:Mparametric}
\ee
with $\alpha_m=10$, $\beta_m=1.5$ and a transition in sign at $z_T=1$ \citep[eqn.\ 20 of][]{ar08}.

The bias surface for this parameterisation (Figure~\ref{fig:mcsys} bottom panel) is well-fit by a plane
$b/\sigma\approx0.031+110m_0-7.5\times10^{10}a_0$.
This means that, while the error budget
\be
|m_0|+6.7\times10^{8}|a_0|\simlt2.8\times10^{-3}
\ee
must again be split between additive and multiplicative systematics, the allocations can still be traded linearly against each other.
Note that absolute requirements on parametric variables $a_0$ and $m_0$ are tighter than those on $\sigc$ and $m$
partly because the unnormalised functions are much lower than unity, and partly because $\add$ and $\mult C$ are now more similar to $\partial C/\partial w$.

\subsection{General forms of additive and multiplicative shear measurement bias} \label{sec:amgeneral}

Since the real scale-dependence of systematics will remain unknown for any survey (even after its completion), we now use a Monte Carlo approach \citep[\cf][]{kit09} to explore {\em all possible} functional forms of $\add$ and $\mult$.
We explore this very high dimensional parameter space separately for each type of bias, but remember the caveat about duplicated error budgets and the necessity/ability to trade between requirements on each.
In general, requirements will emerge upon the functional forms of $\add$ and $\mult$. For tractability, we collapse each function to a single number
\be\label{eqn:sigAdef}
\overline{\add}\equiv
\frac{\sum_{z\cdot\mathrm{bins}} \frac{1}{2\pi} \int_{\ell_{\mathrm{min}}}^{\ell_{\mathrm{max}}}~|\add(\ell,z_A,z_B)|~\ell^2~\mathrm{d}\ln\ell}
{\sum_{z\cdot\mathrm{bins}} \frac{1}{2\pi} \int_{\ell_{\mathrm{min}}}^{\ell_{\mathrm{max}}}~\ell^2~\mathrm{d}\ln\ell}
\ee
\be\label{eqn:sigMdef}
\overline{\mult}\equiv
\frac{\sum_{z\cdot\mathrm{bins}} \frac{1}{2\pi} \int_{\ell_{\mathrm{min}}}^{\ell_{\mathrm{max}}}~|\mult(\ell,z_A,z_B)|~\ell^2~\mathrm{d}\ln\ell}
{\sum_{z\cdot\mathrm{bins}} \frac{1}{2\pi} \int_{\ell_{\mathrm{min}}}^{\ell_{\mathrm{max}}}~\ell^2~\mathrm{d}\ln\ell}.
\ee
Thus we generalise $\sigma^2_{\mathrm{sys}}$ in \citet{ar08} to 3D correlation functions, and include a renormalisation, by way of the denominator, that reduces sensitivity to changes in the adopted $\ell$-range.
Values of these performance indicators are shown on the right and upper axes of Figure~\ref{fig:mcsys}.
For our baseline survey, the denominator in (\ref{eqn:sigAdef}) and (\ref{eqn:sigMdef}) is $55\times9.0$$\times$$10^5$. For the shorter $\ell$-range used by GREAT10, the denominator is $1.8\times10^5$.
Other possible choices for the weighting inside the integral, and the slightly different approach required for practical calculations in GREAT10, are discussed in Appendix~B.

To span the space of possible systematics functions, we generate $100,000$ random realisations of $\add(\ell,z_A,z_B)$; for now, we set $\mult\equiv0$
Assuming conservatively that generic systematics contribute equally to all scales and redshift bins, we generate random systematics by drawing the value of $\sigma_{\bc}(\ell,z)$ in each $\ell$ and $z$ bin from a Gaussian PDF centered about 0.
The width of the Gaussian remains fixed as a function of $\ell$ and $z$, and we repeat this process several times with increasingly wide Gaussians (spanning a range that includes current performance and future requirements).
We then smooth $\sigma_{\bc}$ with a 2D boxcar of width 50 (of 200) $\ell$ bins and 3 (of 10) $z$ bins, and construct $\add(\ell,z_A,z_B)\equiv\sigma_{\bc}(\ell,z_A)\sigma_{\bc}(\ell,z_B)$.
The smoothing reflects the typically continuous form of systematic effects;
it is important here because (unrealistic) realisations of systematics that are uncorrelated between adjacent bins cause less bias in cosmological parameters.
The precise amount of smoothing (particularly in the $\ell$ direction) affects requirements on $\overline{\add}$ by around $15\%$ of the nominal value.
While this precision is adequate for current planning purposes, detailed analysis in the future will require more accurately constrained forms of $\add$ and $\mult$ to be propagated.

\begin{figure}
\includegraphics[width = 8.5cm]{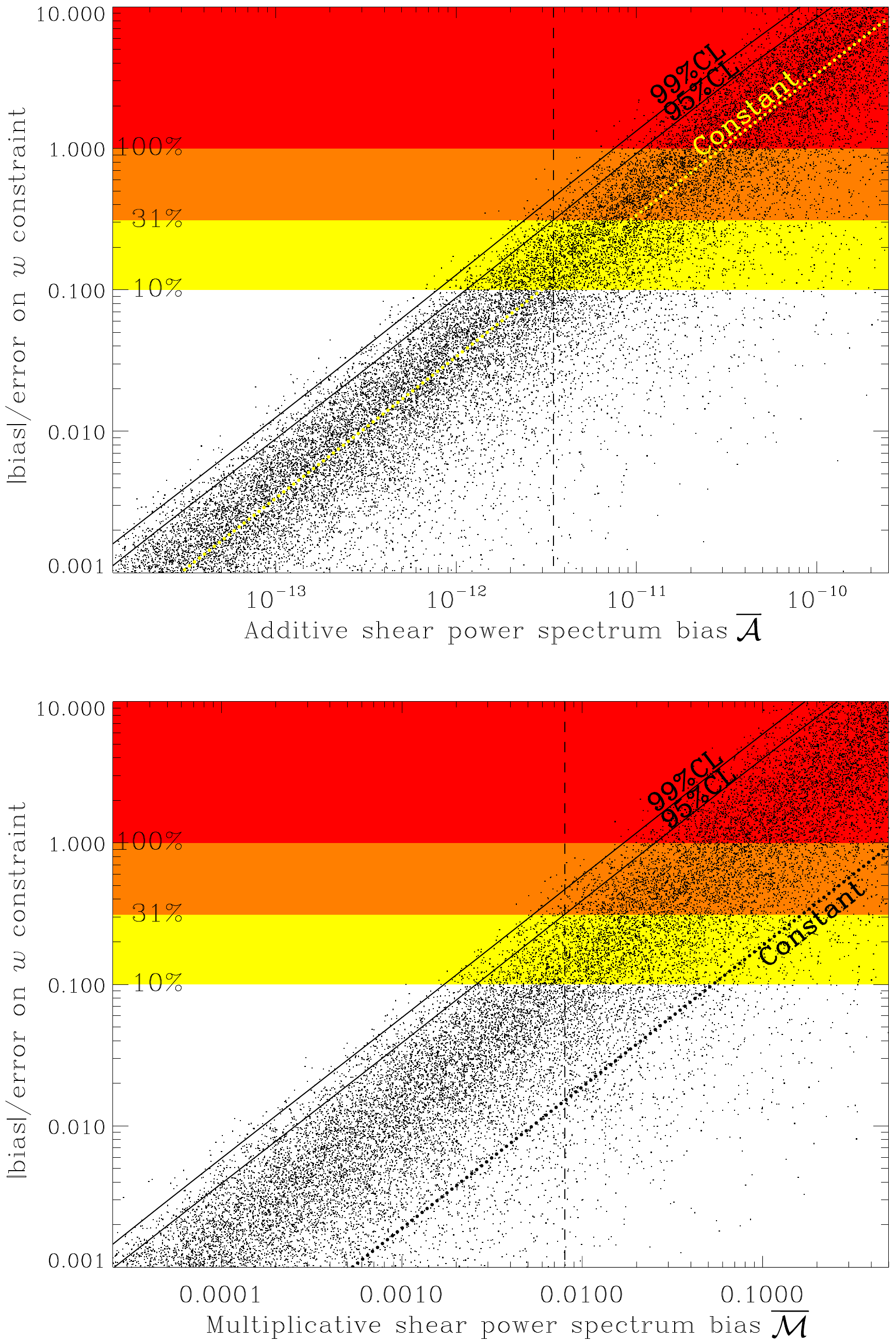}
\caption{\label{fig:asys}
The bias on measurements of the dark energy equation of state parameter $w$ from weak lensing surveys with 
{\bf (top panel)} additive and {\bf (bottom panel)} multiplicative shear measurement systematics.
Each data point shows a random realisation of systematics with a unique dependence upon angular scale and redshift (for clarity, only one in three are plotted).
The dotted diagonal lines show the bias on cosmological parameters if the shear measurement systematics are constant.
The solid diagonal curves show limiting values that include $95\%$ and $99\%$ of random realisations with a given value of $\overline\add$ or $\overline\mult$.
An all-sky 3D cosmic shear survey will only be deemed successful if the measurement bias is $\simlt31\%$ of the statistical measurement error.
At $95\%$CL, this will require (vertical dashed line) shear measurement better than $\overline\add\simlt 3.5\times10^{-12}$ if $\mult\equiv0$,
and $\overline\mult\simlt 8.0\times10^{-3}$ if $\add\equiv0$.}
\end{figure}

We propagate our random realisations of biases on the cosmic shear power spectrum into biases on $w$ using the Fisher matrix bias formalism as before.
The largest biases are generated when the shape of $\add(\ell,z_A,z_B)$ is close to that of $\partial C(\ell,z_A,z_B)/\partial w$.
To ensure that the bias on $w$ is less than $31\%$ of the statistical error for $95\%$ of the random realisations, we require
\be \label{eqn:creq}
\overline\add\simlt 1.8\times10^{-12} 
\ee
(see Figure~\ref{fig:asys}a), including a factor of $1/2$ for a non-zero budget on $\mult$.
This general requirement is a factor of only $\sim3$ tighter than the requirement if $\add$ is constant (see the upper panel of figure~\ref{fig:mcsys} or the dotted line in figure~\ref{fig:asys}a), demonstrating how bad constant additive systematics can be. 
Conversely, it is a factor of $\sim3$ looser than if $\add$ is restricted to the family of curves parameterised by equation~(\ref{eqn:Aparametric}) (see lower panel of figure~\ref{fig:mcsys}), which was a pathological case in the worst $1\%$ of random configurations.

For the smallest resolved galaxies $\ro\approx1.25\rp$ (\ie\ $\rg=0.75\rp$), in the regime of the most elliptical PSF 
typically obtained from astronomical instruments $|\ep|\approx0.1$, and with an Airy PSF such that $\pr\sim2$, equations~\eqref{eqn:sigsdef2_exp} and \eqref{eqn:creq} together become
\ba \label{eqn:clast}
\overline\add \approx 
       0.79\sigma^2{[|\bd\ep|]} 
    + 5.2\sigma^2{[|\bd\ec|]} ~~~~~~~~~~~ \nn
    + \,0.0023 \left\{ \frac{(\delta(\rp^2)^2)}{\rp^4} + \frac{\sigma^2[\rp^2]}{\rp^4} \right\} ~~~~~  \nn
    + \,0.0091 \left\{ \frac{(\delta\rc)^2}{\rc^2} + \frac{\sigma^2[\rc]}{\rp^2} \right\}  ~~~~~~~~~\, \nn
    + \,0.0023~ \frac{(\delta(\ro^2)^2)}{\ro^4} 
    ~~~~~~~   \simlt 1.8\times10^{-12},
\ea
Note that $\bd\ep$ at least is likely to have two components that each contribute to the total bias.

We then generate 100,000 random multiplicative shear measurement biases $m(\ell,z)$ in the same way and with the same smoothing.
We propagate these into multiplicative cosmic shear systematics ${\mult}(\ell,z_A,z_B)$ via equation~(\ref{eqn:Mfromm}), and hence into biases on $w$.
To ensure measurement bias is less than $31\%$ of statistical errors for $95\%$ of the Monte Carlo realisations, we require
\be \label{eqn:mreq}
\overline\mult\simlt 4.0\times10^{-3}
\ee
(see Figure~\ref{fig:asys}b), including a factor of $1/2$ for a finite error budget on $\add$.
This is a factor of $\sim20$ tighter than the requirements if $\mult$ is constant (see the lower panel of figure~\ref{fig:mcsys} or the dotted line in figure~\ref{fig:asys}b), demonstrating again that a constant multiplicative shear measurement bias has surprisingly little effect on $w$ constraints (note that it does strongly affect constraints on $\Omega_m$ and $\sigma_8$).
The amount by which the random systematics are smoothed (particularly in the $z$ direction) affects requirements on $\overline{\mult}$ by around $10\%$ of the nominal value.
For the smallest resolved galaxies, equations~\eqref{eqn:mdef_exp} and \eqref{eqn:mreq} become
\ba \label{eqn:mlast}
\overline\mult \approx
        1.8\frac{\left\langle\delta(\rp^2)\right\rangle}{\left\langle\rp^2\right\rangle} + 
        3.6\left\langle\frac{\delta\rc}{\ro}\right\rangle  + 2\langle\mu\rangle
        \simlt 4.0\times10^{-3},
\ea
plus redundant variance terms that are already constrained more tightly by equation~\eqref{eqn:clast}, so which we drop here.

A top-down analysis can now allocate error budgets to each of the components of $\overline\add$ and $\overline\mult$, as expanded in equations~\eqref{eqn:clast} and \eqref{eqn:mlast}.
In the absence of other information, a natural choice would perhaps allocate budgets in, perhaps in inverse proportion to the coefficient by which they affect the overall science.
\citet{cropper12} provide one such breakdown of these error budgets that is feasible in a dedicated space mission.

\subsection{Comparison to other work} \label{sec:ar08}

Our calculations differ from those of \citet{ar08}, \citet{cha12} and Cardone et al.\ (in prep.) by using a form-filling approach to consider any possible $\ell$-dependence of systematics, rather than just parametric forms.
\citet{ar08} also assumed only a 2D cosmic shear analysis, with a slightly lower redshift distribution of source galaxies, and considered power spectrum measurements up to scales $\ell<20,000$; we exclude such non-linear scales because poorly-understood effects of baryonic physics are likely to make them difficult to interpret \citep{sem11,kit11}.
The denominator we introduced in equations~\eqref{eqn:sigAdef} and \eqref{eqn:sigMdef} keeps our new $\overline{\add}$ and $\overline{\mult}$ performance indicators independent (within a few percent) of this choice of $\ell$-range.
However, if future understanding of small-scale baryonic effects could indeed extend cosmic shear measurements to $\ell=20,000$, statistical errors $\sigma$ would shrink by $\sim$$10\%$.
Exploiting this new information would require correspondingly smaller shear measurement biases.

We can mimic the 2D notation of \citet{ar08} 
by multiplying our 3D requirement~\eqref{eqn:creq} by its denominator, dividing it by the number of (in our case 55) redshift bin pairs that we considered, and including small corrections.
This process, plus small corrections for a few other differences ($\ell$-range, $z$-distribution, $|b|/\sigma$$<$$1$, 100\% CL) yields a pseudo-2D requirement a factor of $\sim$2 looser than their $\sigs^2\simlt 10^{-7}$ per redshift bin.
That difference presumably arises from the details of the redshift slicing, and we shall not consider it further.

\section{Can the requirements be met?}\label{sec:met}

\subsection{Current best shear measurement performance}\label{sec:metcurrent}

The performance of shape measurement algorithms can be tested on simulated astronomical images that contain a known shear signal.
Blind competitions include the community-wide STEP \citep{step1,step2} and GREAT \citep{great08handbook,great10} programmes;
these have been and are continuing to be supplemented by efforts by individual groups targeted towards specific surveys.
Assessed using the GREAT metric Q, these programmes have yielded a steady improvement by a factor $\sim$3.5 per year over the past decade \citep{kaggle}.

GREAT10 is the most recent blind competition, and the first to employ variable shear simulations, which are required to test scale-dependent issues.
The best methods entered into GREAT10 achieve
$\overline\add\sim2.7\times10^{-12}$ and $\overline\mult\sim3.1\times10^{-3}$ on bulge+disc galaxies at detection S/N=40 \citep[Table~4 of ][in which these values are expressed as $\sqrt{\overline{\add}}$ and $\overline{\mult}/2$, but see Appendix~B for a discussion of slight differences in approach]{g10res}. 
For these fairly bright galaxies, current performance surpasses the requirement on $\overline{\mult}$ and the requirements on $\overline{\add}$ and $\overline{\mult}$ can be traded against each other to also be met in combination.
Note, however, that this shape measurement inaccuracy uses all but $1\%$ of the entire error budget.
GREAT10 assumed a spatially/temporally varying PSF\footnote{The GREAT10 simulations used ground-based PSF morphologies, but STEP3 (see \url{http://www.roe.ac.uk/~heymans/step/step3_results.html}) concluded that the only factor affecting shear measurement performance was the ratio of the PSF size to the pixel size. STEP3 was a space-based equivalent of STEP2, run as another public, blind competition. Its results were never published because they were essentially identical to those from STEP2.The main conclusion was that equivalent shear measurement performance could obtained for small galaxies from space as for similarly-resolved larger galaxies from the ground, irrespective of PSF morphology.}, but that it was perfectly known, and that non-convolution effects could be perfectly corrected.
Further development in shape measurement will be necessary if part of the error budget is to be set aside for \egg\ PSF or CTI modelling errors.

Faint galaxies are harder to measure, but must be included to reach Stage~IV surveys' statistical goals on cosmological parameter estimation.
At detection S/N=20, the best methods now achieve $\overline\add\sim2.1\times10^{-11}$ and $\overline\mult\sim5.6\times10^{-3}$; at detection S/N=10 they achieve $\overline\add\sim7.4\times10^{-11}$ and $\overline\mult\sim1.1\times10^{-2}$.
If all galaxies were this faint, exploiting them (consuming all of the available error budget) would exceed requirements in $\sqrt{\overline{\add}}$ by a factor $3.5$--$6.5$ and in $\overline{\mult}$ by a factor $1.4$--$2.8$.
If an analysis were to proceed using extant shear measurement methods, accounting for residual systematic biases would necessarily enlarge the reported error bars --- if all galaxies were at detection S/N=10, 95\% of realisations of bias would simultaneously satisfy $|b|/\sigma$$<3.7$ for $\overline{\add}$ and $|b|/\sigma$$<1$ for $\overline{\mult}$  (see figure~\ref{fig:asys}).

Shape measurement algorithms can be improved either by fundamental progress or by calibration on accurate simulated images. 
Extrapolating the current rate of fundamental development \citep{kaggle} suggests that, with even minimal continued development, the required algorithmic performance will be surpassed, and substantial margin will be achieved, well before the need to analyse Stage~IV surveys.
Indeed, noise bias \citep{kac12,mel12} was unaccounted for by all GREAT10 methods, but appears in faint galaxies at a level consistent with its being the dominant source of bias \citep{ref12}.
Proper treatment of noise bias will therefore improve performance for faint galaxies.
Several additional improvements have also been suggested \citep[\egg][]{ber10,vio11}.
For the first time, methods are thus emerging with sufficient accuracy to reliably and fully exploit the statistical potential of Stage~IV cosmic shear surveys.
Simulations could then be used solely as external verification tests of data analysis pipelines.
Dedicated simulation efforts are continuing inside the teams of all weak lensing surveys\footnote{See \url{www.darkenergysurvey.org},  \url{www.lsst.com}, \url{www.euclid-ec.org}, \url{http://www.naoj.org/Projects/HSC/}, \citet{halo}.}, and the GREAT3 programme \citep{great3} is currently being designed by a worldwide collaboration of the weak lensing community.

\subsection{Empirical diagnosis of residual additive systematics}

Although the greatest improvement is formally required in additive cosmic shear measurement biases, they are potentially the least troublesome.
Many additive systematics can be internally diagnosed within a shear catalogue, and those that do arise can potentially even be calibrated out at the catalogue level.
This procedure has a long heritage in Hubble Space Telescope (HST) analyses \citep[\egg][]{rho04,sch04,rho07, jee07,sch10,jee11,hoe11}.

\subsubsection{Calibrating PSF model errors} \label{sec:addpsf}

The best way to internally diagnose PSF modelling errors $\delta\rp$ and $\bd\ep$ is to bootstrap real stellar shapes. 
The PSF model can be constructed from all but a few of the available stars, 
then interpolated to the positions and colours of the remaining stars as well as the galaxies. 
Any offset between the predicted and measured values will be a sum of $\delta\rp+\delta\ro$ and $\bd\ep+\bd\eo$, 
but the observational contributions should average to zero over a large population of stars.
The number of degrees of freedom in PSF variation due to thermomechanical instability \citep{JB05,rho07, sch10}, atmospheric turbulence \citep{jar09} or changing gravity load \citep{subaru} can also be usefully compared to engineering predictions from raytracing through optics models \citep{tinytim1,tinytim2}.

A vital test of successful PSF deconvolution is obtained from the correlation of measured shears with the PSF ellipticity.
No residual signature of the system's PSF should find its way into the galaxy shape catalogue, so these should be uncorrelated. 
However, in a flawed shear measurement, taking unweighted $\widehat{\ep}\equiv\ep+\bd\ep$ and $\widehat{\boldsymbol{\gamma}}_{\rm w}$ from equations~\eqref{eqn:stepclast} and \eqref{eqn:stepmlast}, we obtain
\ba
\left\langle\widehat{\boldsymbol{\gamma}}_{\rm w}.\widehat{\ep}\right\rangle=
 \frac{\variance{\ep}}{\pg\pr\pep} \frac{\rp^2}{\rg^2}
  \Bigg(\frac{\delta(\rp^2)}{\rp^2} + 
      \frac{2\,\delta\rc}{\ro-\rc} ~~ \hspace{-20mm} \nn
   + \frac{\delta(\ro^2)}{\ro(\ro-\rc)} 
   + \frac{\pr\rg^4}{\rp^2(\pr\rg^2+\rp^2)}\frac{\delta\pr}{\pr} \Bigg) \hspace{-20mm} \nn
   -  \, \frac{1}{\pg\pr} \frac{\rp^2}{\rg^2} \left( \frac{\left\langle\bd\ep\right\rangle^2+\sigma^2[|\ep|]}{\pep} \right)~~~~~~~~~~~~~~~~ \hspace{-20mm} \nn
   + \left\langle{\boldsymbol{\gamma}}_{\rm w}.\bd{\ep}\right\rangle,~~~~~~~~~~~~~~~~~~~~~~~~~~~~~~~~~~~~~~~~~~~~~~~~~ \hspace{-20mm}
\ea
plus many more terms of order $\mathcal{O}(\delta^2)$, including some proportional to equations~\eqref{eqn:crossterm4} and \eqref{eqn:crossterm5}.
While it would be difficult to identify and then calibrate out any individual contribution from this mixed observable,
it can be used as an invaluable {\em post facto} check that other techniques have successfully removed almost all of the additive cosmic shear systematics.

\subsubsection{Calibrating residual detector effects}

Non-convolution detector effects can accumulate in space-based instruments over time, as radiation damages the hardware. 
Thus any long-term, monotonic drift in the mean $\langle\ro\rangle$ or $\langle\eo\rangle$ within each exposure 
-- or, even better, within a calibration field that can be returned to -- indicates a nonzero $\delta\rc$ or $\bd\ec$.

Many detector effects also exhibit a characteristic dependence upon chip position.
This is most notable for Charge Transfer Inefficiency in CCDs, where the image degradation increases linearly with distance $y$ from the readout register \citep{cti1},
where $y_{\mathrm{max}}$ is the size of the CCD.
In this case, correlating shear measurements with chip position, or fitting shear measurements as a function of chip position, measures nonzero
\ba \label{eqn:meang}
 \Big\langle \boldsymbol{\gamma}_{\rm w} \Big\rangle \Big|_{y_{\mathrm{max}}} =
 \frac{1}{\pg\pr\pec}\frac{\pr\rg^2+\rp^2}{\rg^2}\bd\ec \big|_{y_{\mathrm{max}}}  ~~~~~ \nn
- \frac{2\mean{\ep}}{\pg\pr\pep} \frac{\rp^2}{\rg^2} \frac{\delta\rc}{\ro} \Bigg|_{y_{\mathrm{max}}} ~~~~~~~~~~~~~~~ \nn
 - \frac{1}{\pg\pr\pep}\frac{\rp^2}{\rg^2} \Bigg\{ \mean{\bd\ep} + ~~~~~~~~~~~~~ \nn
 \mean{\ep} \Bigg( \frac{\delta(\rp^2)}{\rp^2} + \frac{\delta(\ro^2)}{\ro(\ro-\rc)} + ~~~~ \nn
 \frac{\pr\rg^4}{\rp^2(\pr\rg^2+\rp^2)}\frac{\delta\pr}{\pr} \Bigg) \Bigg\} 
\ea
where we assume $\bd\ec|_{y_{\mathrm{max}}}$ and $\delta\rc|_{y_{\mathrm{max}}}$ are constant over a sufficiently long time period to gather statistically significant measurements.
If $\mean{\ep}=0$ and all other (PSF, observational) errors were zero, this would be a direct test of $\bd\ec$.
However, the reality that \eqref{eqn:meang} contains terms mixed with residual PSF modelling errors has made analysis of HST data challenging.
Only by first verifying the PSF model with tests from Section~\ref{sec:addpsf}, 
\citet{rho07,sch10,hoe11} were able to subtract this measurement of $\bd\ec$ from a shear catalogue, following equation~\eqref{eqn:egdef_rjm}.
However, such an empirical, catalogue-level correction should be seen as a last resort because it addresses neither $\delta\rc$ nor the mixing
between sources of error whereby an imperfect removal of additive systematics can introduces an
(undiagnosable) multiplicative cosmic shear systematic.
A much more robust technique, demonstrated by \citet{lea10}, is to apply a physically-motivated 
correction scheme at the pixel level as the first process during data reduction \citep[\egg][]{cti3,jay10}.
The performance of this technique can again be tested via equation~\eqref{eqn:meang},
and improved by iteration.

\subsection{Impact of residual multiplicative systematics}

Multiplicative cosmic shear measurement biases are potentially the most troublesome, because there is no known cosmology-independent way to accurately 
diagnose residual multiplicative bias internally within a dataset (except that it may leak weakly into a small unphysical $B$-mode 
signal \citep{val06}, but so do many things). 
Analyses must either rely upon theoretical calculations of the shear calibration,
or test a measurement pipeline on simulated images that contain a known signal
and rely upon the veracity of those simulations.
Since multiplicative systematic errors are thus more problematic than additive errors, and because the requirements on them are similarly hard to meet, we shall investigate them more carefully.

Rather than considering galaxies all of the same size and detection S/N, we shall now consider a realistic, full population of galaxies.
Some galaxies are bigger and brighter than others, and it will be easier to measure their shapes. 
The form of equation~\eqref{eqn:stepmlast} suggests that multiplicative shape measurement biases
predominantly depend upon the relative size of the PSF and the surveyed galaxies
\be \label{eqn:stepgeneric}
m\approx m_0+\frac{m_1}{\pr}\left(\frac{\rp^2}{\rg^2}\right).
\ee
This characteristically quadratic performance was indeed apparent in many of the methods tested in 
STEP2\footnote{In STEP2, methods that applied an overall `calibration factor' from analysis of independent simulated images (\egg\ TS and several not plotted) appear to have achieved $\langle m\rangle\approx0$ by adjusting $m_0$ such that $m(\langle\rg\rangle)=0$ for galaxies of average size.} \citep[][top-right panels of figure~7]{step2}.
Similar behaviour is suggested in GREAT08 \citep[][figure~C3]{great08} and is explicitly fitted in GREAT10 \citep[][appendix~B5]{g10res} as
\be \label{eqn:g10alpha}
m\approx m_0+\alpha_{R^2_{\mathrm{PSF}}}\frac{\langle\rg^2\rangle}{\langle\rp^2\rangle}\frac{\rp^2}{\langle\rg^2\rangle}
\ee
where $\langle\rp^2\rangle=3.4^2$~pixels$^2$, $\langle\rg^2\rangle\approx3.55^2$~pixels$^2$ (averaging the contribution of the bulges and discs), and the best methods achieve $\alpha_{R^2_{\mathrm{PSF}}}\approx0.005$ \citep[][figure~5]{g10res}. 
Note that GREAT10's fiducial PSF had a Moffat profile, for which $\pr\sim1$. Diffraction-limited surveys with $\pr\sim2$ will likely achieve better performance although, since that was only tested in a subset of the GREAT10 data whose results were dominated by method bias, we shall conservatively assume only the performance explicitly demonstrated.

We showed in Section~\ref{sec:amconst} that constraints on the nature of dark energy are largely insensitive to a constant multiplicative bias $m_0$.
The achieved value of $m_1$ is thus likely to be the driving requirement for success.
We shall baseline a currently achievable performance of $m_0\approx0$ and $m_1\approx0.006$.
We shall then fold through the observed distribution of galaxies sizes to
consider the prospects of two generic regimes proposed for future surveys\footnote{A survey's effective $\rp$ may be a complicated function of the system PSF at different times. Some state-of-the-art shear measurement algorithms downweight the contribution from exposures with poor seeing. This improves the effective $\rp$, at a cost of decreased imaging depth.}: 
a space-based mission with a PSF Full Width at Half Maximum (FWHM) of $0\farcs 2$ 
and a ground-based telescope with a FWHM seeing of $0\farcs 7$.

\subsubsection{Two dimensional cosmic shear} \label{sec:2dcs}

\begin{figure}
\includegraphics[width = 8.5cm]{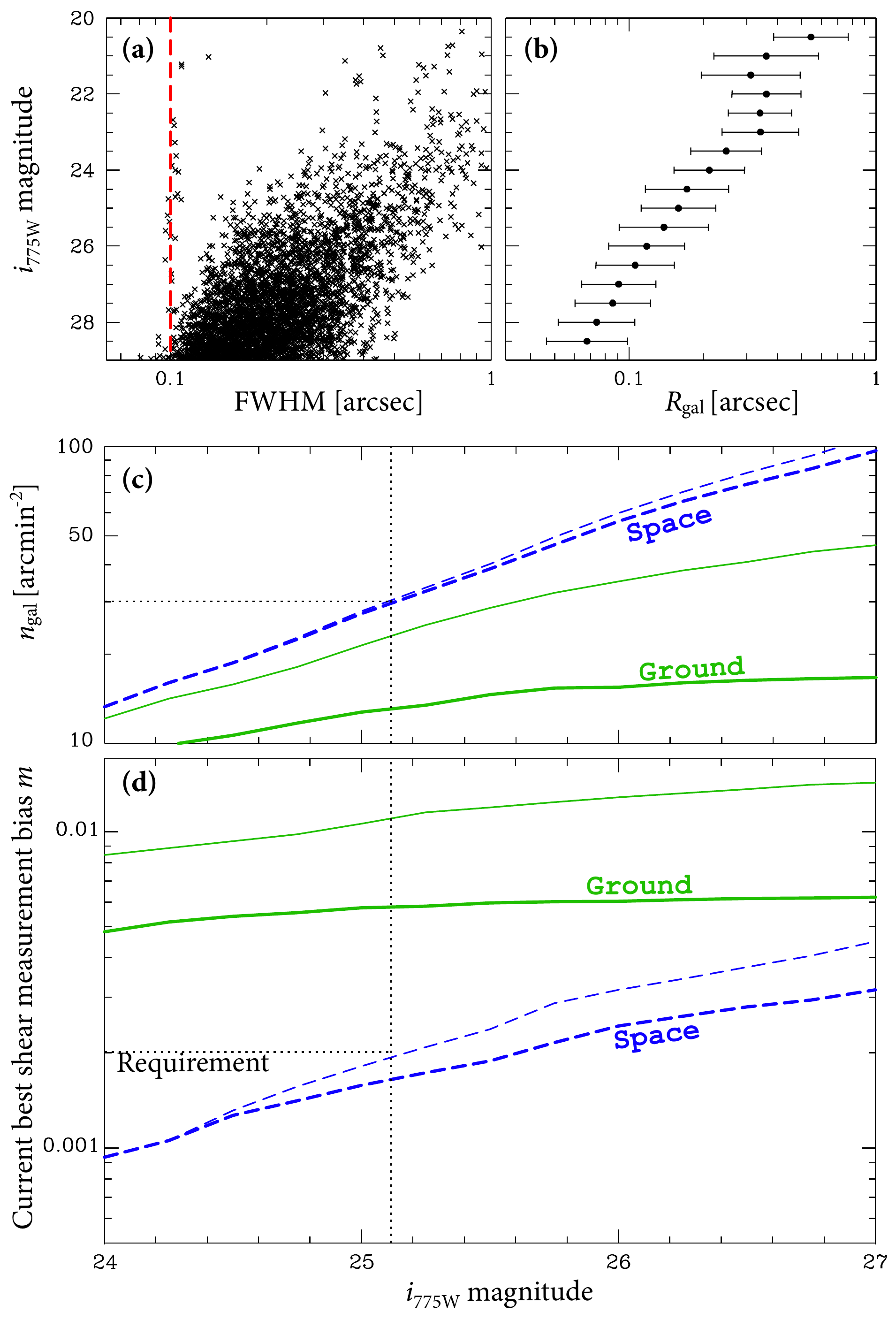}
\caption{\label{fig:udf}
Prospects for 2D weak gravitational lensing surveys.
{\bf Panel (a):} The observed size $\ro$ and $i$-band magnitude of objects in the UDF.
The vertical dashed line indicates the size of the ACS PSF. 
{\bf Panel (b):} Galaxies' average intrinsic size $\rg$ as a function of magnitude, 
under the assumption that the galaxies and PSF have Gaussian profiles. 
The error bars indicate the dispersion in $\rg$.
{\bf Panel (c):} The cumulative number density of resolved galaxies as a function of (limiting) magnitude,
with sizes $\ro>1.25\rp$ (thick lines) or $\ro>1.1\rp$ (thin lines). 
The dashed lines correspond to a space based mission with a FWHM$=0\farcs2$ for the PSF. 
Note that Euclid's wide-band observations to magnitude 24.5 correspond roughly to $i_{\mathrm{775W}}\approx25.2$ (vertical dotted line).
The solid curves are for a typical ground based PSF with FWHM$=0\farcs7$. 
{\bf Panel (d):} Predicted shear measurement bias for the best current methods, 
averaged over the population of resolved galaxies.
Requirement \eqref{eqn:mreq} is shown as a horizontal dotted line, assuming $\overline{\mult}\approx 2m$ (\ref{eqn:Mapprox2m}).}
\end{figure}

To quantify the typical size of galaxies in the Universe as a function of magnitude, 
we measure the sizes of galaxies in $i_{\mathrm{775W}}$-band observations of 
the HST Ultra Deep Field \citep[UDF;][]{udf} (Figure~\ref{fig:udf}a).
To compute the approximate intrinsic size of the galaxies, we assume that their profiles are Gaussian (with a FWHM equal
to their measured FWHM), and that the ACS PSF has a FWHM of $0\farcs1$. 
Fainter galaxies are smaller (Figure~\ref{fig:udf}b) but, down to $i_{\mathrm{775W}}\simlt26$, most
are intrinsically larger than the ACS PSF.

Many more galaxies are resolved $(\ro>1.25\rp)$ by the hypothetical space-based mission 
than the hypothetical ground-based survey (Figure~\ref{fig:udf}c).
Crucially, most galaxies in space-based observations are not only resolved but {\em very well} resolved.
Following~\eqref{eqn:stepgeneric}, this naturally leads to a better shear measurement bias (Figure~\ref{fig:udf}d).
For a full, realistic population of source galaxies in a 2D cosmic shear survey from space, 
current shear measurement performance satisfies requirement \eqref{eqn:mreq}, in the absence of PSF variation or detector effects.
Any subsequent improvement will provide increased margin for imperfect PSF and detector models.

\begin{figure}
 \includegraphics[width = 8.5cm]{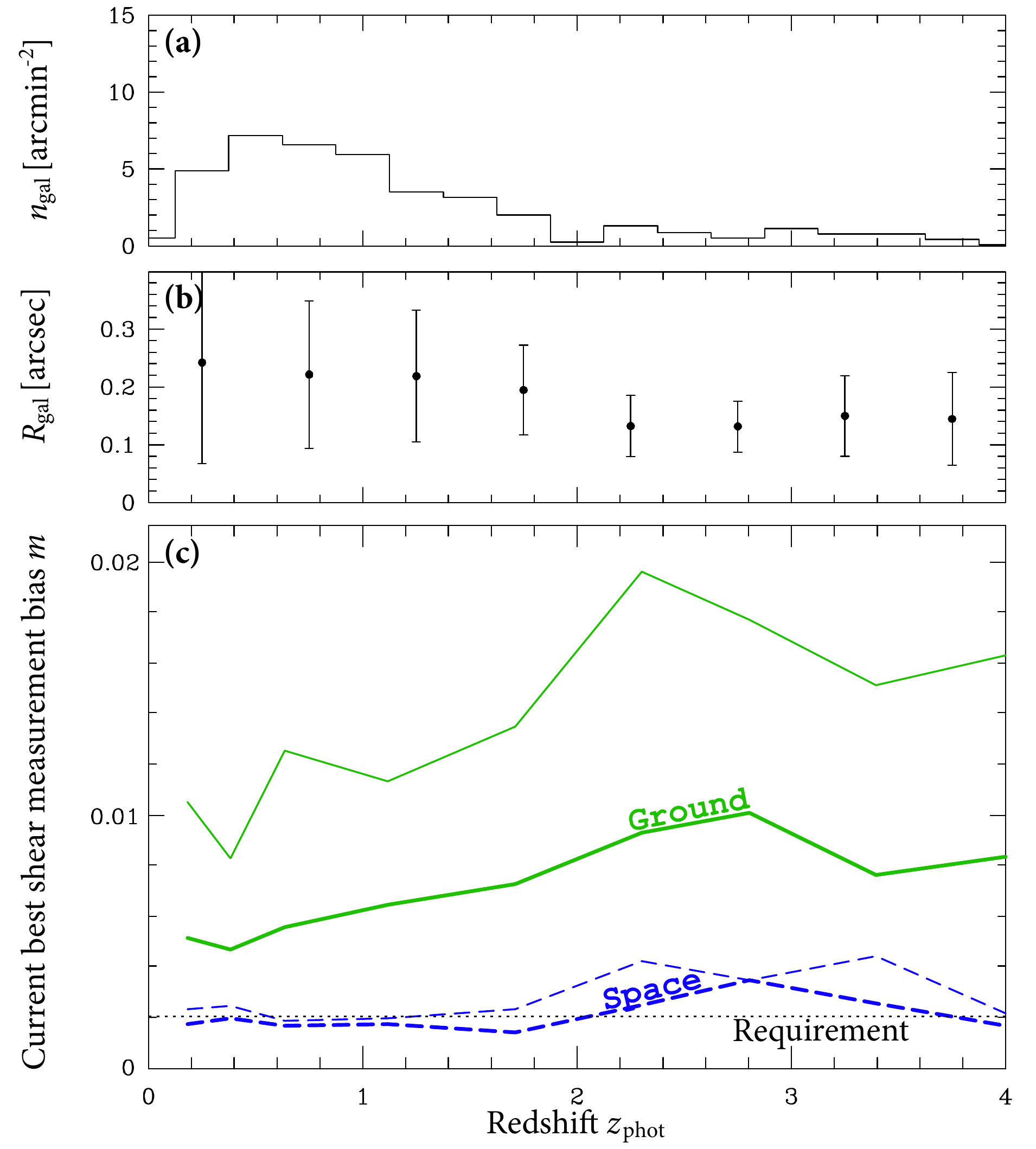}
\caption{\label{fig:z}
Prospects for 3D weak gravitational lensing surveys.
{\bf Panel (a):} The density of galaxies as a function of photometric 
redshift $z_{\rm phot}$ for galaxies with $20<i_{775W}<26.5$
(results do not depend strongly on the choice of limiting magnitude).
{\bf Panel (b):} Average galaxy size $\langle\rg\rangle$ as a function of redshift,
under the assumption that the galaxies have Gaussian profiles. 
The error bars indicate the dispersion in $\rg$.
{\bf Panel (c):} Multiplicative shear calibration bias $m$ as a
function of redshift for galaxies with sizes $\ro>1.25\rp$ (thick lines) or $\ro>1.1\rp$ (thin lines).
The dashed line corresponds to a space-based mission with a PSF of FHWM$=0\farcs2$,
and the solid curve is for typical ground-based seeing with FHWM$=0\farcs7$.
Requirement \eqref{eqn:mreq} is shown as a dotted line, assuming $\overline{\mult}\approx 2m$ (\ref{eqn:Mapprox2m}).}
\end{figure}

Ground-based surveys face two problems.
First, a greater improvement in shape measurement techniques is required for them to reach their full potential than space-based surveys (Figure~\ref{fig:udf}d).
This is simply because of the difficulty {\em resolving} galaxies from the ground, {without even taking into account the much harder task of modelling the PSF due to a turbulent atmosphere and more variable physical conditions}.
Second, even in extremely deep images covering the entire sky, not enough galaxies are resolved $(\ro>1.25\rp)$
to obtain statistical measurement errors on $w$ competitive with other techniques 
(Figure~\ref{fig:udf}c).
More galaxies could be included in an analysis by lowering the size cut\footnote{It is far more effective to add small galaxies than faint ones, 
especially for a ground-based survey, because faint galaxies are also so much smaller.
In practice, our crude step-function cut is also usually replaced by a smoothly varying weight function 
\citep{Hoekstra00}; the result of this will lie between the two extremes we have considered.}, for example to 
$\ro>1.1\rp$.
Increasing the density of galaxies reduces statistical measurement error, 
but at a cost of even more rapidly increasing systematic bias, such that current methods do not meet requirements.

\subsubsection{Three dimensional cosmic shear} \label{sec:3dcs}

Three dimensional cosmic shear analysis requires measurements of both shear and redshift for each galaxy, and for the shears to be measured without (even relative) bias as a function of redshift \citep{kit11}.
To estimate this bias in a real population of galaxies, 
we use photometric redshift estimates for $20<i_{775W}<26.5$ galaxies in the HST UDF by \citet{coe06}.
The distribution of best-fit redshifts peaks 
around $z\sim0.5$ but also samples a long tail out to $z\sim3$ (Figure~\ref{fig:z}a). 
Beyond redshift $z\sim3$, the scarcity of UDF galaxies makes our statistics unstable.

The mean and rms apparent size of galaxies decrease noticeably above $z\sim1.5$--$2$ (Figure~\ref{fig:z}b). 
Multiplicative shear measurement bias will therefore get slightly worse at high redshift (Figure~\ref{fig:z}c).
For a space-based survey, meeting requirements in every redshift bin will demand algorithms with multiplicative biases a factor $1.8$--$2.2$ better than current methods (which could come from calibration on very accurate simulated images).
Note that this analysis is completely independent of that in Section~\ref{sec:metcurrent}.
That their conclusions are so consistent lends support to both methodologies.

Ground-based observations are more profoundly affected by the decrease in galaxy size at $z\simgt1.5$.
Very deep images will help, because some fraction of systematics is doubtless due to noise bias \citep{ref12}.
However, a dramatic improvement in shear measurement methods will be required for ground-based surveys to span the high redshifts needed to probe the growth of structure.
As before, this argument is based purely on the small size of galaxies compared to a ground-based PSF, and does not take into account additional challenge of modelling the more variable ground-based PSF.

\section{Discussion and conclusions} \label{sec:conc}

We have derived expressions showing how various sources of error in galaxy shape measurement 
propagate into additive biases $\add$ (eqn.~\ref{eqn:sigsdef2_exp}) and multiplicative biases $\mult$ (eqn.~\ref{eqn:mdef_exp}) on cosmic shear results.
Additive biases include a contribution from mis-estimation of a telescope's PSF shape, and multiplicative biases include mis-estimation of the PSF size.
This agrees with the behaviour generically seen in empirical tests of shear measurement methods.
For the first time, we have also propagated into cosmic shear results the consequences of imperfect correction for non-linear detector effects, and 
imperfect image processing algorithms.

We have ascertained the maximum level of additive biases $\add(\ell,z)$ (eqn.~\ref{eqn:creq}) and multiplicative biases $\mult(\ell,z)$ (eqn.~\ref{eqn:mreq}) that can be tolerated by a next-generation cosmic shear survey attempting to constrain the dark energy equation of state parameter $w$ to within $0.065$ ($68\%$CL).
Cosmic shear measurements of $w$ are surprisingly insensitive to a {\em constant} multiplicative bias.
To explore more generic scale- and redshift-dependent systematic biases, we have used a form-filling technique;
based upon the $95\%$ confidence limit averaged equally over all possible functional forms, we define convenient requirements on mean $\overline\add$ and $\overline\mult$.
\citet{cropper12} distribute this overall requirement into budgets on the individuals sources of error
(PSF knowledge, detector knowledge, accuracy of shape measurement algorithms) in an allocation that is suitable for a real space mission.

We compare our requirements on galaxy shape measurement software to the performance seen recently in the public, blind GREAT10 challenge.
Extant shear measurement methods meet both requirements for a Stage~IV weak lensing surveys, for bright galaxies at detection S/N=40 or for a 2D cosmic shear survey from space in which the contributions from a large population of galaxies are combined.
This will generally not provide sufficient galaxies to meet Stage~IV surveys' goals for the statistical errors on cosmological parameters.
This also assumes that the telescope and instrument hardware can be well modelled; a modest improvement will create margin for imperfect modelling and correction of the system PSF or detector effects.

Fully exploiting the statistical potential of Stage~IV weak lensing surveys will require shear measurement software that works more accurately than current algorithms on faint galaxies.
Current algorithms could introduce systematic biases of the same order of magnitude as the statistical errors, and the total reported confidence limits would need to be enlarged by a factor $\sim$$\sqrt{2}$ to account for this effect.
To be sure of avoiding this problem, if all galaxies were detected at S/N=20--10 and all of them were used, additive biases must be reduced by a factor 3.6--6.5.
However, many tests can be used to identify and remove portions of a shear catalogue with additive biases; we have used our new formalism to show exactly what each test is sensitive to.
Using an entire, realistic population of faint galaxies would also need a reduction in multiplicative bias by a factor 1.4--2.8.
Averaging over a realistic galaxy population extending to $z\simgt1.5$, a space-based 3D cosmic shear analysis will need an improvement in multiplicative bias by a factor 1.8--2.2.
No internal tests can identify multiplicative biases, so the greatest development effort should be spent to minimise these.

Several new ideas for image analysis techniques are being discussed in the literature, and ongoing simulation programmes show potential.
The past decade has seen steady improvement in shape measurement algorithms; extrapolating even minimal continued development suggests that the required algorithmic performance will be met well before the need to analyse Stage~IV surveys.
Importantly, it will be at least 3-5 times easier to meet requirements for high resolution space-based rather than ground-based surveys, because multiplicative biases depend (theoretically and empirically) on the inverse square of the S/N and the square of the PSF size.
This conclusion that ground-based surveys will require much better shear measurement methods than space-based surveys arises solely because they do not resolve galaxies well, and does not even take into account the additional challenge of modelling atmospheric turbulence or more rapidly changing physical conditions.

\vspace{-2mm}
\section*{Acknowledgments}

RM and TK are supported by Royal Society University Research Fellowships.
HH is supported by the Netherlands Organization for Scientific Research through VIDI grants and acknowledges support from the Netherlands Research School for Astronomy (NOVA). 
RM and HH also acknowledge support from ERC International Reintegration Grants.
This work was done in part at JPL, run under a contract for NASA by Caltech.
TS is supported by the NSF through grant AST-0444059-001, and by the Smithsonian Astrophysics Observatory through grant GO0-11147A.
YM acknowledges support from  CNES and CNRS/INSU

\noindent {\it Facilities:} This paper uses data from observations with the Hubble Space Telescope HST-GO-9978 (P.I.\ S.~Beckwith).

\section*{Appendix A: Remaining cross terms} \label{sec:crossterms}

In Section~\ref{sec:problems}, we ignored several cross terms in earlier calculations of the additive cosmic shear systematic $\add$ because we expect their contributions to be subdominant as long as the PSF model, detector characterisation and shape measurement method are working properly.
However, tests for the presence of these terms in real data could be a useful, cosmology-independent way to verify that the pipeline is meeting requirements. 
We shall now discuss four noteworthy order $\mathcal{O}(\delta)$ terms that potentially add to $\add$. These are
\ba \label{eqn:crossterm1}
-\left\langle\eg.\ep\right\rangle
\left\langle\frac{\rp^2}{\rg^2}\right\rangle
\times ~~~~~~~~~~~~~~~~~~~~~~~~~~~~~~~ \nn
       \left(\frac{\delta(\rp^2)}{\rp^2} +
      \frac{2\delta\rc}{\ro-\rc} +
  \frac{\delta(\ro^2)}{\rg^2+\rp^2}\right)
\ea
in the presence of the selection bias discussed by \citet{hs03}, whereby galaxies are more likely to be detected if their intrinsic shapes are similar to that of the PSF;
\be \label{eqn:crossterm2}
-\left\langle\eg.\bd\eo\right\rangle
\left\langle\frac{\rg^2+\rp^2}{\rp^2}\right\rangle
\ee
if, for example, a faulty shape measurement method systematically truncates the isophotes of elliptical galaxies; and
\be \label{eqn:crossterm3}
+\left\langle\eg.\bd\ec\right\rangle
\left\langle\frac{\rg^2+\rp^2}{\rp^2}\right\rangle
\ee
with Charge Transfer Inefficiency, for which $\bd\ec$ depends on $\eg$ \citep{rho10}; and
\be \label{eqn:crossterm4}
-\left\langle\eg.\bd\ep\right\rangle
\left\langle\frac{\rp^2}{\rg^2}\right\rangle
\ee
if some small galaxies (which have been sheared, so correlate with their neighbours) are accidentally confused with stars and allowed to contribute towards the PSF model. Of all these, the first two terms of \eqref{eqn:crossterm1} are likely to be the most problematic: the first because stars and galaxies have different colours, so a PSF model na\"ively obtained from stars will be systematically too large, and the second because model inaccuracies in nonlinear correction will likely dominate variations in the effect across the detector.

There are also several terms of order $\mathcal{O}(\delta^2)$. 
Two that may feasibly have nonzero coefficients are
\be \label{eqn:crossterm5}
+\left\langle\bd\ep.\frac{\delta\rp^2}{\rp^2}\ep\right\rangle
\left\langle\frac{\rp^4}{\rg^4}\right\rangle
\ee
if the PSF modelling errors depend upon the ellipticity of a complex PSF whose shape changes as a function of radius; and
\be \label{eqn:crossterm6}
+\left\langle\bd\ep.\bd\ec\right\rangle
\left\langle\frac{\rp^2(\rp^2+\rg^2)}{\rg^4}\right\rangle
\ee
if residuals from the correction of nonlinear detector effects also contaminate the bright stars from which the PSF is modelled.

Finally, we also ignored cross terms like $\langle m\bmath{\gamma}\bc \rangle$ in the correlation functions.
Ideally, $\langle m\bmath{\gamma}\bc \rangle=\langle m\bc \rangle\,\langle\bmath{\gamma}\rangle$ and $\langle\bmath{\gamma}\rangle=0$, but this latter equality does not hold in the presence of \citet{hs03} selection biases.
Furthermore, we have shown that $m$ and $\bc $ are both correlated with $\delta\rp$ and therefore with each other, so the prefactor may be considerable.
This sort of combination could give rise to a whole new slew of potential intrinsic-intrinsic, intrinsic-$\bc $, intrinsic-$m$, etc.\ systematics.
We shall explore these in future work.

\section*{Appendix B: Performance indicators used in GREAT10} \label{sec:great10}

In equations~\eqref{eqn:sigAdef} and \eqref{eqn:sigMdef}, we introduced performance indicators $\overline{\add}$ and $\overline{\mult}$, based upon integrals over a range of scales.
For consistency with earlier work \citep{ar08}, we chose to weight the scales by $\ell^2~\mathrm{d}\ln\ell$, but different choices could have been made.
Integration with respect to $\mathrm{d}\ell$ typically raises the numerical value of $\overline{\add}$ by $\sim10\%$ and $\mult$ by $\sim3\%$.
A similar loosening would also need to be applied to the numerical value of the requirements, and this is a negligible change.
Including weighting by $C(\ell)\mathrm{d}\ell$ inside the integrals in \eqref{eqn:sigMdef} rescales the performance indicator and requirement so that they have a numerical value similar to $\overline{\add}$.
However, this would mean losing intuition from previous studies and make the requirements formally cosmology-dependent.
Furthermore, since the shape of the $C(\ell)$ weight approximately recovers that of $\ell^2~\mathrm{d}\ln\ell$, changes to numerical values are even smaller than the previous option.

Practical considerations forced the measurements in GREAT10 \citep{g10res} to use a different range in $\ell$ and a different weight function.
It is important to consider the effect of this, because we use the GREAT10 results as an indication of current best performance.
The GREAT10 analysis measured $\widehat{C}(\ell)$ at linearly separated values $\ell=\{233$, $415$, $600$, $789$, $977$, $1162$, $1350$, $1538\}$, then found the least-squares fitting function $(1+{\mult})C(\ell)+{\add}$ with constant $\add=\overline{\add}_{\mathrm{G}}$ and $\mult=\overline{\mult}_{\mathrm{G}}$.
This process thus minimises
\be
\chi^2({\add},{\mult}) \equiv \sum_\ell \left(\widehat{C}(\ell) - ( 1+{\mult})C(\ell) -{\add} \right)^2.
\ee
Therefore
\be
\frac{\partial\chi^2}{\partial{\add}}=2\sum_\ell \left(\widehat{C}(\ell) - (1+{\mult})C(\ell) -{\add} \right)=0
\ee
so, if ${\mult}=0$,
\be
\sum_\ell{\add} = \sum_\ell \left(\widehat{C}(\ell)-C(\ell) \right).
\ee
Approximating the discrete sums with constant $\Delta\ell$ as continuous integrals, and remembering that  $\add=\overline{\add}_{\mathrm{G}}$ is constant so can be extracted from the integrals,
\be
\overline{\add}_{\mathrm{G}} = \frac{\frac{1}{2\pi}\int_{\ell_{\mathrm{min}}}^{\ell_{\mathrm{max}}}(\widehat{C}(\ell)-C(\ell))~\mathrm{d}\ell}
 {\frac{1}{2\pi}\int_{\ell_{\mathrm{min}}}^{\ell_{\mathrm{max}}} ~\mathrm{d}\ell}.
\ee
This is similar to equation~\eqref{eqn:sigAdef}, although a version in which the various $\ell$ scales are weighted differently.
The different weighting changes our conclusions by less than $10\%$, so we ignore this small perturbation.

Least-squares fitting also guarantees that
\be
\frac{\partial\chi^2}{\partial{\mult}}=-2\sum_\ell C(\ell)\left(\widehat{C}(\ell) - (1+{\mult})C(\ell) -{\add} \right)=0
\ee
so, if ${\add}=0$,
\ba
\overline{\mult}_{\mathrm{G}}=\frac{\sum C(\ell)(\widehat{C}(\ell)-C(\ell))}{\sum \left(C(\ell)\right)^2} ~~~~~~~~~~~~~\\
 =\frac{\frac{1}{2\pi}\int_{\ell_{\mathrm{min}}}^{\ell_{\mathrm{max}}}C(\ell)(\widehat{C}(\ell)-C(\ell))~\mathrm{d}\ell}
 {\frac{1}{2\pi}\int_{\ell_{\mathrm{min}}}^{\ell_{\mathrm{max}}} (C(\ell))^2 ~\mathrm{d}\ell}.
\ea
This again is merely a differently-weighted version of equation~\eqref{eqn:sigMdef}, with negligible effect upon our conclusions.

\bsp
\label{lastpage}

\end{document}